\documentclass[iop]{emulateapj-rtx4}
\usepackage{mathrsfs }
\usepackage{natbib}
\usepackage{amssymb}
\usepackage{ulem} 
\newcommand{\secpoint}{\mbox{$''\mskip-7.6mu.\,$}}

\def\kms{km~s$^{-1}$}
\def\cm2{cm$^{-2}$}

\def\ltsima{$\; \buildrel < \over \sim \;$}
\def\gtsima{$\; \buildrel > \over \sim \;$}
\def\simgt{\lower.5ex\hbox{\gtsima}}
\def\simlt{\lower.5ex\hbox{\ltsima}}

\usepackage{float}

\begin{document}

\title{The MOSDEF Survey: Excitation Properties of $\lowercase{z}\sim 2.3$ Star-forming Galaxies\altaffilmark{1}}
\author{
 Alice E. Shapley,\altaffilmark{2}
 Naveen A. Reddy,\altaffilmark{3,4} 
 Mariska Kriek,\altaffilmark{5}
 William R. Freeman,\altaffilmark{3}
 Ryan L. Sanders,\altaffilmark{2} 
 Brian Siana,\altaffilmark{3} 
 Alison L. Coil,\altaffilmark{6} 
 Bahram Mobasher,\altaffilmark{3}
 Irene Shivaei,\altaffilmark{3} 
 Sedona H. Price,\altaffilmark{5} \&
 Laura de Groot,\altaffilmark{3}
 }

\altaffiltext{1}{Based on data obtained at the W.M. Keck Observatory, which is operated as a scientific partnership among the California Institute of Technology, the University of California,  and the National Aeronautics and Space Administration, and was made possible by the generous financial support of the W.M. Keck Foundation.}
\altaffiltext{2}{Department of Physics and Astronomy, University of California, Los Angeles, 430 Portola Plaza, Los Angeles, CA 90095, USA}
\altaffiltext{3}{Department of Physics and Astronomy, University of California, Riverside, 900 University Avenue, Riverside, CA, 92521, USA}
\altaffiltext{4}{Alfred P. Sloan Fellow}
\altaffiltext{5}{Astronomy Department, University of California at Berkeley, Berkeley, CA 94720, USA}
\altaffiltext{6}{Center for Astrophysics and Space Sciences, Department of Physics, University of California, San Diego, 9500 Gilman Drive., La Jolla, CA 92093, USA}

\email{aes@astro.ucla.edu}


\shortauthors{Shapley et al.}


\shorttitle{MOSDEF: Excitation Properties at $z\sim 2.3$}

\begin{abstract}
We present results on the excitation properties of $z\sim2.3$ galaxies using early
observations from the MOSFIRE Deep Evolution Field (MOSDEF) Survey. With its coverage
of the full suite of strong rest-frame optical emission lines,
MOSDEF provides an unprecedented view of the rest-frame optical
spectra of a representative sample of distant star-forming galaxies. We investigate the locations 
of $z\sim 2.3$ MOSDEF galaxies in multiple emission-line diagnostic diagrams. These
include the [OIII]$\lambda 5007$/H$\beta$ vs. [NII]/H$\alpha$ and [OIII]$\lambda5007$/H$\beta$
vs. [SII]$\lambda\lambda6717,6731$/H$\alpha$ ``BPT" diagrams, as well as the
$O_{32}$ vs. $R_{23}$ excitation diagram. We recover
the well-known offset in the star-forming sequence of high-redshift
galaxies in the [OIII]$\lambda 5007$/H$\beta$ vs. [NII]/H$\alpha$ BPT diagram relative to Sloan Digital Sky Survey
star-forming galaxies. However, the shift for our rest-frame optically
selected sample is less significant
than for rest-frame-UV selected and emission-line selected galaxies at $z\sim 2$.
Furthermore, we find that the offset is mass-dependent, only appearing within
the low-mass half of the $z\sim 2.3$ MOSDEF sample, where galaxies are shifted
towards higher [NII]/H$\alpha$ at fixed [OIII]/H$\beta$. Within the [OIII]$\lambda5007$/H$\beta$
vs. [SII]/H$\alpha$ and $O_{32}$ vs. $R_{23}$
diagrams, we find that $z\sim 2.3$ galaxies are distributed like local ones,
and therefore attribute the shift in the [OIII]$\lambda 5007$/H$\beta$ vs. [NII]/H$\alpha$ BPT diagram to elevated N/O abundance
ratios among lower-mass ($M_*<10^{10} M_{\odot}$) high-redshift galaxies.
The variation in N/O ratios calls into question the use at high
redshift of oxygen abundance indicators based on nitrogen lines, but the apparent invariance
with redshift of the excitation sequence in the $O_{32}$ vs. $R_{23}$ diagram paves the way
for using the combination of $O_{32}$ and $R_{23}$ as an unbiased metallicity
indicator over a wide range in redshift. This indicator
will allow for an accurate characterization of the
shape and normalization of the mass-metallicity relationship over more than 10 Gyr.
\end{abstract}

\keywords{galaxies: evolution --- galaxies:high-redshift --- galaxies: ISM}

\section{Introduction}
\label{sec:intro}

Rest-frame optical emission-line spectra contain a wealth of information
about the fundamental properties of galaxies. These include the instantaneous
star-formation rate, the degree of nebular dust extinction, the electron
density in star-forming regions, the gas-phase chemical abundance, and the nature
of the ionizing energy source, be it hot stars, an active galactic nucleus
(AGN), or shocks. Accordingly, the nebular emission lines in the spectra of both individual H~II regions
and the integrated light from galaxies follow well-defined patterns
and correlations that reflect the range of underlying physical conditions in 
star-forming regions and active nuclei.  These patterns have been well-traced
among the ensemble of the strongest rest-frame optical emission lines, including
[OII]$\lambda\lambda3726,3729$, H$\beta$, [OIII]$\lambda\lambda 4959,5007$, H$\alpha$,
[NII]$\lambda6584$, and [SII]$\lambda\lambda6717,6731$.

Building on earlier work by \citet{baldwin1981}, \citet{veilleux1987}
proposed three diagnostic emission-line diagrams featuring the following
axes: [OIII]$\lambda 5007$/H$\beta$ vs. [NII]$\lambda 6585$/H$\alpha$,
[OIII]$\lambda 5007$/H$\beta$ vs. [SII]$\lambda\lambda 6717,6731$/H$\alpha$,
and [OIII]$\lambda 5007$/H$\beta$ vs. [OI]$\lambda 6300$/H$\alpha$, commonly
referred to as ``BPT" diagrams in reference to the original \citeauthor{baldwin1981} work. These
diagrams can be used to distinguish the ionizing mechanism in emission-line galaxies, 
specifically whether it is from hot stars or an active nucleus.
Star-forming galaxies occupy very well-defined loci in these diagnostic
diagrams, in particular in the [OIII]$\lambda 5007$/H$\beta$ vs. [NII]$\lambda 6585$/H$\alpha$ diagram.
As metallicity increases, the sequence of star-forming galaxies
in the space of [OIII]$\lambda 5007$/H$\beta$ vs. [NII]$\lambda 6585$/H$\alpha$ extends from high values of [OIII]$\lambda 5007$/H$\beta$ and 
low [NII]$\lambda 6585$/H$\alpha$, and curves down to low [OIII]$\lambda 5007$/H$\beta$
and high [NII]$\lambda 6585$/H$\alpha$. Galaxy stellar mass also increases
along the sequence, due to the relationship between stellar mass and gas-phase
metallicity in star-forming galaxies \citep{tremonti2004}.
Another diagnostic diagram commonly used to characterize star-forming galaxies
is the space of [OIII]$\lambda\lambda$4959,5007/[OII]$\lambda\lambda$3726,3729 ($O_{32}$) 
vs. ([OIII]$\lambda\lambda$4959,5007+[OII]$\lambda\lambda$3726,3729)/H$\beta$ ($R_{23}$). 
Photoionization models \citep{ferland1998,kewley2002,dopita2013}
tuned to match the distributions of local galaxies in these diagnostic diagrams have been
used to translate observed sets of emission lines into physical quantities
such as oxygen abundance, ionization parameter  (i.e., ratio of ionizing photon to particle density
in H~II regions), and hardness of the ionizing spectrum.

Over the past 15 years, rest-frame optical measurements have been assembled for
galaxies at $z>1$ using near-IR spectrographs on large ground-based telescopes, 
providing a window into distant star-forming regions.
Based on typically small samples and incomplete sets of rest-frame
optical emission lines (e.g., [NII]/H$\alpha$ in the K-band at $z\sim 2$,
or H-band at $z\sim 1.5$), we have begun to probe the evolution in galaxy metallicity
\citep[e.g.,][]{erb2006a}, dust content \citep[e.g.,][]{dominguez2013,price2014}, and H~II region physical
conditions \citep[e.g.,][]{shirazi2014}. 

Intriguingly, early observations of H$\beta$, [OIII]$\lambda5007$, H$\alpha$,
and [NII]$\lambda6584$ suggested a systematic offset in the [OIII]$\lambda 5007$/H$\beta$ vs. [NII]$\lambda 6585$/H$\alpha$ star-forming sequence
of galaxies at $z>1$, relative to that of local galaxies in the Sloan Digital
Sky Survey (SDSS) \citep{shapley2005b,liu2008,erb2006c}. These observations
have inspired many different explanations, including evolution with redshift
in the typical interstellar pressure, and typical
H~II-region ionization parameters, electron densities, density
structure, and ionizing spectra
\citep{liu2008,brinchmann2008,kewley2013,yeh2013}. Alternatively,
the offset has been explained as the contribution by weak, unresolved AGN emission
\citep[e.g.,][]{wright2010}. Furthermore, \citet{juneau2014} have highlighted
the importance of accounting for selection effects at high redshift. 
The BPT offset also raised red flags about inferring
oxygen abundances in high-redshift galaxies using 
locally-calibrated strong-line metallicity indicators \citep{pp2004}, which are tied to either
the ratio of [NII]$\lambda6584$/H$\alpha$ (the $N2$ indicator) or that of 
([OIII]$\lambda5007$/H$\beta$)/([NII]$\lambda6584$/H$\alpha$) (the $O3N2$ indicator).
Indeed, if the high-redshift galaxies follow different patterns from those of local
galaxies in the [OIII]$\lambda 5007$/H$\beta$ vs. [NII]$\lambda 6585$/H$\alpha$ BPT diagram, it seems likely that a different translation is required between
strong-line emission ratios and oxygen abundance \citep{liu2008,newman2014}.

Despite signaling some potentially important physical changes in galaxies
at high redshift, the samples used to measure their emission-line excitation properties 
to date have been small and limited to the [OIII]$\lambda 5007$/H$\beta$ vs. [NII]$\lambda 6585$/H$\alpha$ diagnostic diagram. To assemble the full
picture, we need a statistical and unbiased sample of galaxies with coverage across multiple diagnostic
diagrams.  Now with the advent of multi-object near-IR spectrographs on $8-10$-meter-class
telescopes, we are poised to systematically characterize the emission-line
patterns among high-redshift galaxies, using the full suite of strong rest-frame
optical emission lines spanning in wavelength from [OII]$\lambda\lambda3726,3729$
to [SII]$\lambda6717,6731$. The MOSFIRE Deep Evolution Field (MOSDEF) survey
provides the ideal dataset for such investigations. When complete,
MOSDEF will contain rest-frame optical spectra of $\sim 1500$ galaxies 
at $z\sim 1.4-3.8$, including sensitive observations of all of the relevant 
strong rest-frame optical emission lines at $z\sim 1.6-2.6$, and most features over
the full redshift range. Furthermore, as MOSDEF is conducted in {\it Hubble
Space Telescope} ({\it HST}) legacy fields \citep{grogin2011,koekemoer2011}, each MOSDEF target
has extensive multi-wavelength ancillary data. Therefore, rest-frame optical
emission-line properties can be connected with other basic galaxy properties
such as stellar mass, star-formation history, and morphology. In this paper, we
demonstrate the power of the MOSDEF survey for characterizing the physical properties
of star-forming galaxies at $z\sim 2$, based on their patterns of rest-frame
optical emission lines.

In Section~\ref{sec:observations}, we give a brief overview of the MOSDEF
survey, describing our observations, measurements, and sample selection. Section~\ref{sec:results}
presents our results on the excitation diagrams of $z\sim 2.3$ MOSDEF galaxies, and how
they relate to those in the local universe. In Section~\ref{sec:discussion}, we conclude by discussing
the implications of the excitation properties of $z\sim 2.3$ galaxies for abundance ratio
variations, and proposing a new oxygen abundance indicator for high-redshift galaxies.
Throughout this paper we assume a $\Lambda$-CDM cosmology with $H_{0} = 
70$ \kms\  Mpc$^{-1}$, $\Omega_{\rm m} = 0.3$, and $\Omega_{\Lambda} = 0.7$.

\section{Observations and Analysis}
\label{sec:observations}

\subsection{The MOSDEF Survey}
\label{sec:observations-mosdef}
Our analysis is based on early observations from the MOSDEF
survey. MOSDEF is a four-year project using the MOSFIRE
spectrograph \citep{mclean2012} on the 10~m Keck~I telescope to measure
the physical properties of galaxies at $1.4 \leq z \leq 3.8$.
The full details of the MOSDEF survey design, observations, and analysis
are described in \citet{kriek2014}.
In brief, we target galaxies over 500 square arcminutes in the 
AEGIS, COSMOS, and GOODS-N
extragalactic legacy fields with {\it Hubble Space Telescope} ({\it HST}) imaging
coverage from the Cosmic Assembly Near-infrared Deep Extragalactic Legacy Survey \citep[CANDELS;][]{grogin2011,koekemoer2011} and HST/WFC3 grism spectroscopy
from the 3D-HST survey \citep{brammer2012}.  All MOSDEF targets have extensive 
multi-wavelength ancillary data and robust photometric redshifts, while
$\sim 40$\% have previously-determined spectroscopic redshifts. MOSDEF
targets fall in three redshift intervals ($1.37 \leq z \leq 1.70$,
$2.09 \leq z \leq 2.61$, and $2.95 \leq z \leq 3.80$), which are dictated
by the desire to cover rest-frame optical emission lines within windows
of atmospheric transmission, and are selected down to limiting {\it HST}/WFC3 F160W magnitudes of 24.0,
24.5, and 25.0, respectively, at $z\sim 1.5, 2.3$, and $3.4$. When complete, MOSDEF will include rest-frame
optical spectra for $\sim 1500$ galaxies, with roughly half the sample at $2.09 \leq z \leq 2.61$,
and the other half evenly divided between the other two redshift intervals.

\subsection{Observations, Reductions, and Analysis}
\label{sec:observations-observations}
In this paper, we focus on galaxies at $z\sim 2.3$, which comprise the bulk of the
sample in early MOSDEF observations. For these galaxies, we collected J, H, and K-band spectra,
to maximize the set of strong optical emission lines covered at rest-frame
$3700 - 7000$~\AA. The nominal exposure time in each band is 2 hours per filter,
corresponding to 3$\sigma$ H$\alpha$ emission-line detections for objects with unobscured
star-formation rates of $\sim 1 \mbox{ M}_{\odot} \mbox{ yr}^{-1}$.
The data presented here were collected over the course of
five observing runs from 2012 December to 2013 May and include eight
MOSFIRE masks.\footnote{In addition to data collected on observing runs specifically scheduled
for the MOSDEF project, H- and K-band observations were obtained by K. Kulas, I. McLean,
and G. Mace in 2013 May for one MOSDEF mask in the GOODS-N field.} In addition to the main
MOSDEF target fields, we obtained a single mask in each of the GOODS-S and UDS fields, due
to the lack of field visibility in 2012 December. Each mask had $\sim 30$ 0\secpoint7 slits, 
yielding a resolution of, respectively, 3300 in J, 3650 in H, and 3600 in K.
For galaxies at $z\sim2.3$, we typically cover [OII]$\lambda\lambda$3726,3729
in the J band, H$\beta$ and [OIII]$\lambda\lambda$4959,5007 in the H band,
and H$\alpha$, [NII]$\lambda$6584, and [SII]$\lambda\lambda$6717,6731
in the K band.

We reduced the data in two dimensions using a custom IDL pipeline \citep[see][for a full description]{kriek2014},
yielding both signal and error frames for each slit. 
One-dimensional signal and error spectra for both primary targets
and serendipitous objects falling on each slit were then optimally extracted
(Freeman et al., in prep.). The relative spectral response was established with
observations of B8 -- A1 V telluric standard stars, and tied to an absolute scale
by matching the flux density in the spectrum of a reference star on the
mask to its broadband photometry. Such reference stars were drawn from the 3D-HST
photometric catalogs, with H$_{AB} \leq 20.5$. As described in detail in \citet{kriek2014},
this initial calibration was refined for each
galaxy based on the predicted amount of differential slit loss relative to the reference
star, which is a function of {\it HST} morphology and seeing. Our careful methodology for
flux calibration is required to estimate emission-line ratios spanning multiple filters, with data
potentially collected under different seeing conditions in different filters.  As a test of 
this method, we compared the measured spectroscopic flux density and the flux density in the best-fit spectral
energy distribution to the multi-wavelength photometric dataset for 
the sample of objects with detected continuum. The overall agreement between these two
sets of flux densities suggests no significant
bias in the slit-loss corrected line fluxes, with the assumption that line emission
has the same spatial distribution as that of the continuum. This analysis also suggests that the
uncertainty from slit
loss corrections for line ratios spanning multiple filters is $\approx 18$\% \citep{kriek2014}.

We measured emission-line fluxes with Gaussian line profile fits to the extracted, 
flux-calibrated one-dimensional spectra. The highest signal-to-noise (S/N) emission
line (typically H$\alpha$ or [OIII]$\lambda 5007$) was used to obtain an initial estimate
of the nebular redshift and FWHM. The corresponding values for all other detected
emission lines were constrained to be close to these initial estimates \citep{kriek2014}. Furthermore,
the [OII] and [SII] doublets were each fit as the sum of two Gaussians, while the combination of H$\alpha$ and the [NII]
doublet was deblended with three Gaussians. Line-flux uncertainties were estimated
through Monte Carlo simulations in which we perturbed the one-dimensional spectra
a large number of times according to their error spectra, re-measuring emission lines 
in each perturbed spectrum, and measuring the standard deviation of the resulting
distribution of perturbed fluxes. The MOSFIRE redshift for each galaxy, which is used for
our analysis, was estimated from the observed centroid of the highest S/N emission line.
Balmer emission-line fluxes were corrected for underlying stellar absorption based
on the equivalent widths of stellar Balmer features as estimated from the stellar population
synthesis model fit to the multi-wavelength spectral energy distribution (SED)
of each galaxy (Reddy et al., in prep). 
Balmer absorption is negligible for H$\alpha$, but manifests as a median correction of 14, 12, and 6\%,
for H$\beta$, respectively, within the [OIII]$\lambda 5007$/H$\beta$ vs. [NII]$\lambda 6585$/H$\alpha$, 
[OIII]$\lambda5007$/H$\beta$ vs. [SII]/H$\alpha$, and $O_{32}$ vs. $R_{23}$ samples 
described in Section~\ref{sec:observations-sample}.

\begin{figure}[b!]
\includegraphics[width=9cm]{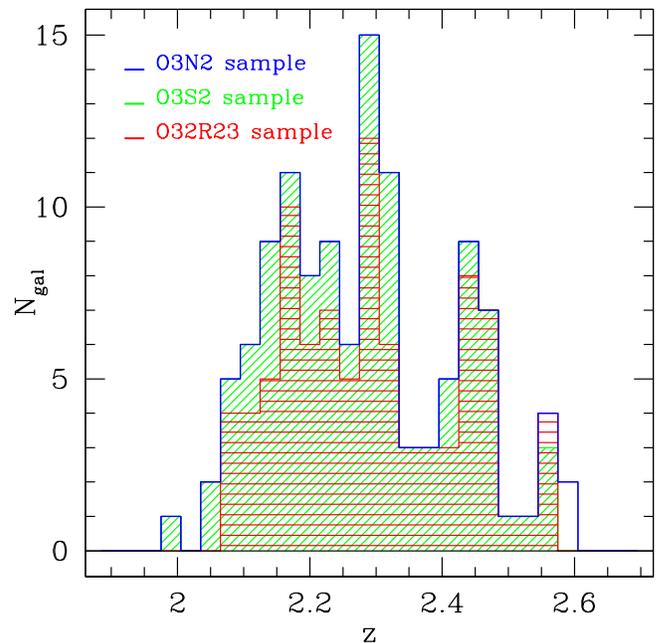}
\caption{MOSFIRE redshift histograms for MOSDEF galaxies with excitation measurements. The blue histogram
indicates the sample of 118 galaxies with coverage of H$\beta$, [OIII]$\lambda 5007$,
H$\alpha$, and [NII]$\lambda6584$ (the [OIII]$\lambda 5007$/H$\beta$ vs. [NII]$\lambda 6585$/H$\alpha$ sample). The green histogram shows the sample of
115 galaxies with coverage of  H$\beta$, [OIII]$\lambda 5007$, H$\alpha$, and
[SII]$\lambda\lambda6717,6731$ (the  [OIII]$\lambda5007$/H$\beta$
vs. [SII]/H$\alpha$ sample).  The red histogram denotes the sample of 89
galaxies with coverage of [OII]$\lambda\lambda3726,3729$, H$\beta$, [OIII]$\lambda 5007$, and H$\alpha$
(the $O_{32}$ vs. $R_{23}$ sample). The average redshift for all three samples is $\langle z \rangle=2.3$.}
\label{fig:zhist_samples}
\end{figure}

\begin{figure*}[t!]
\centerline{\includegraphics[height=1.35in]{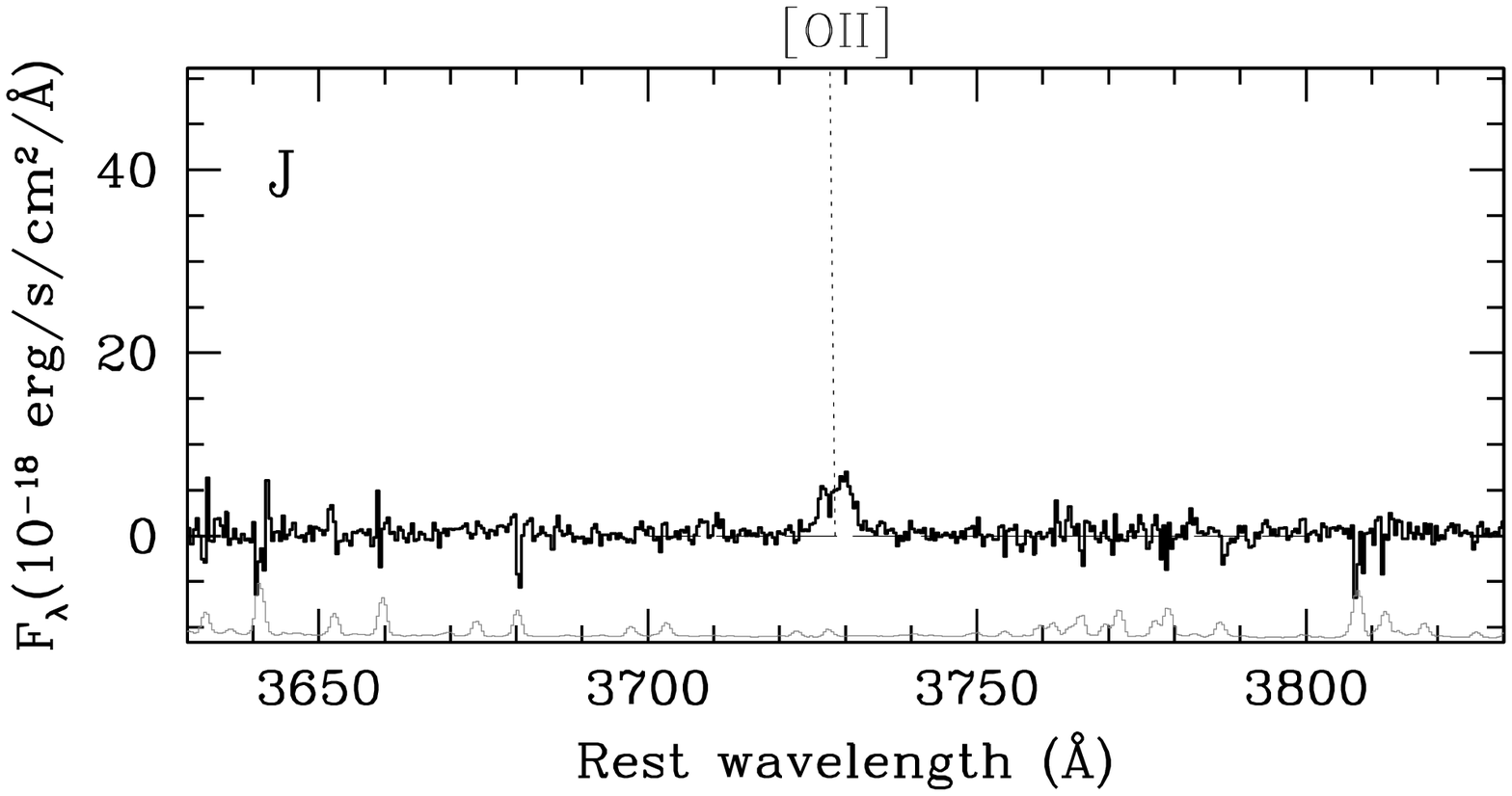}\includegraphics[height=1.35in]{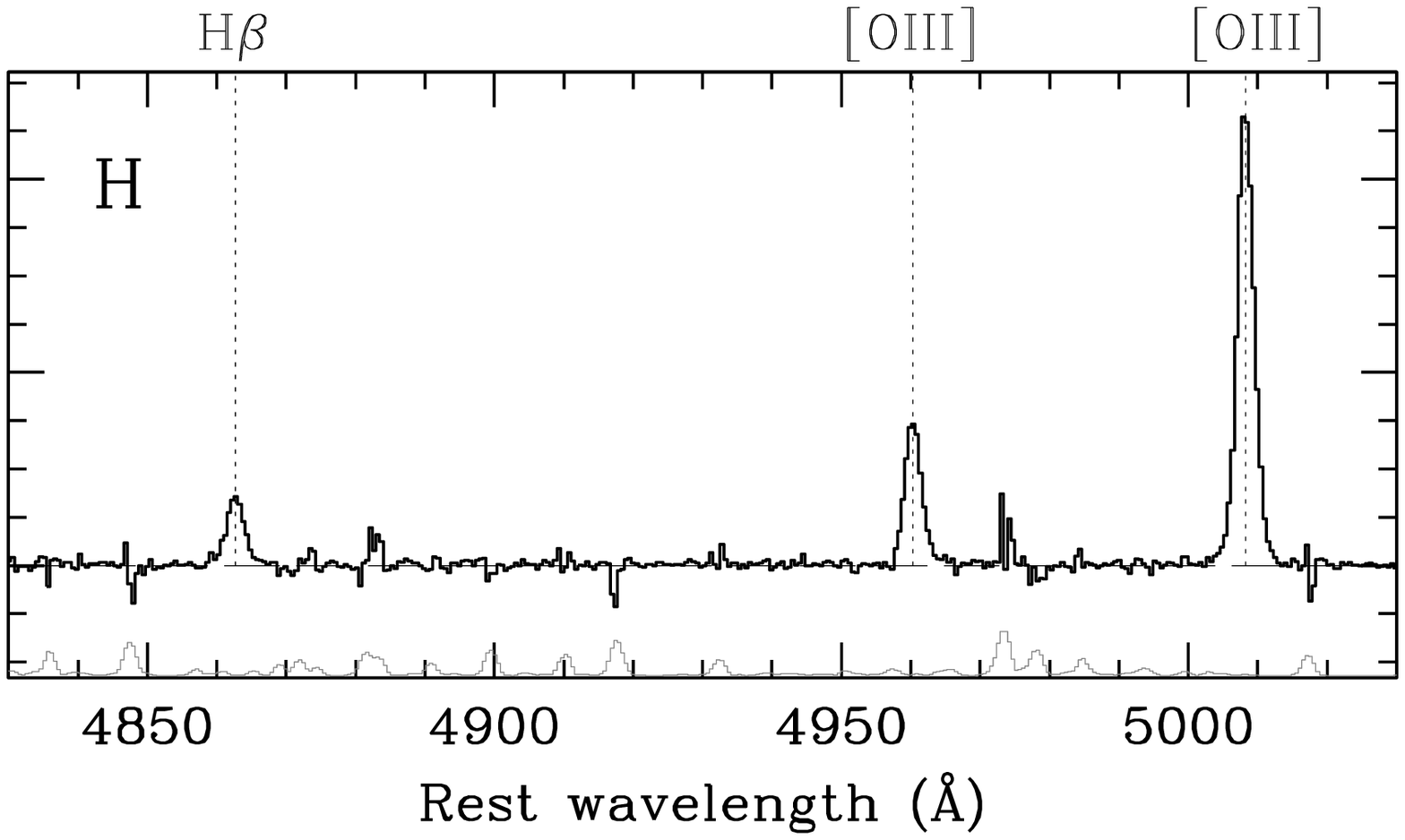}\includegraphics[height=1.35in]{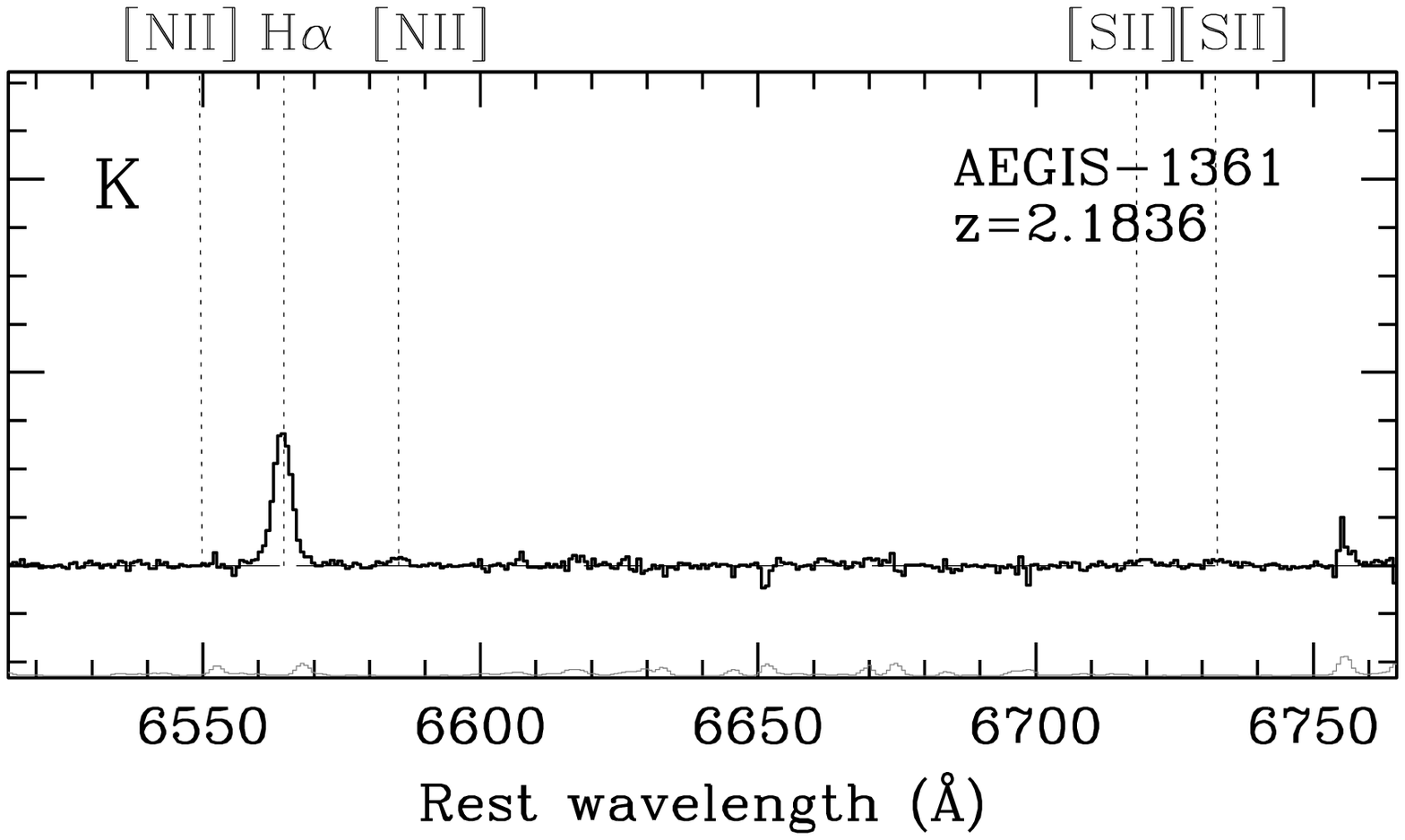}}
\centerline{\includegraphics[height=1.35in]{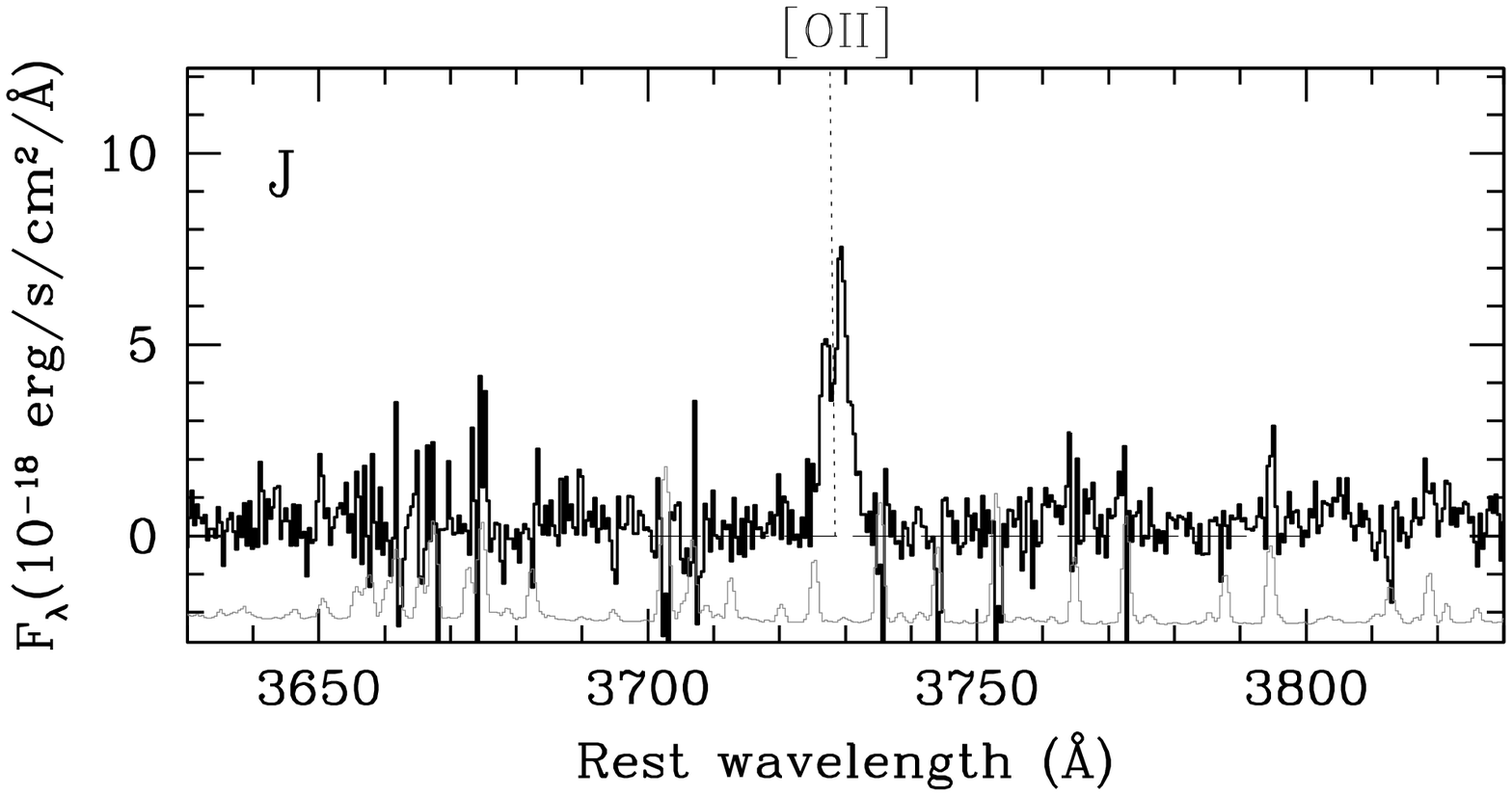}\includegraphics[height=1.35in]{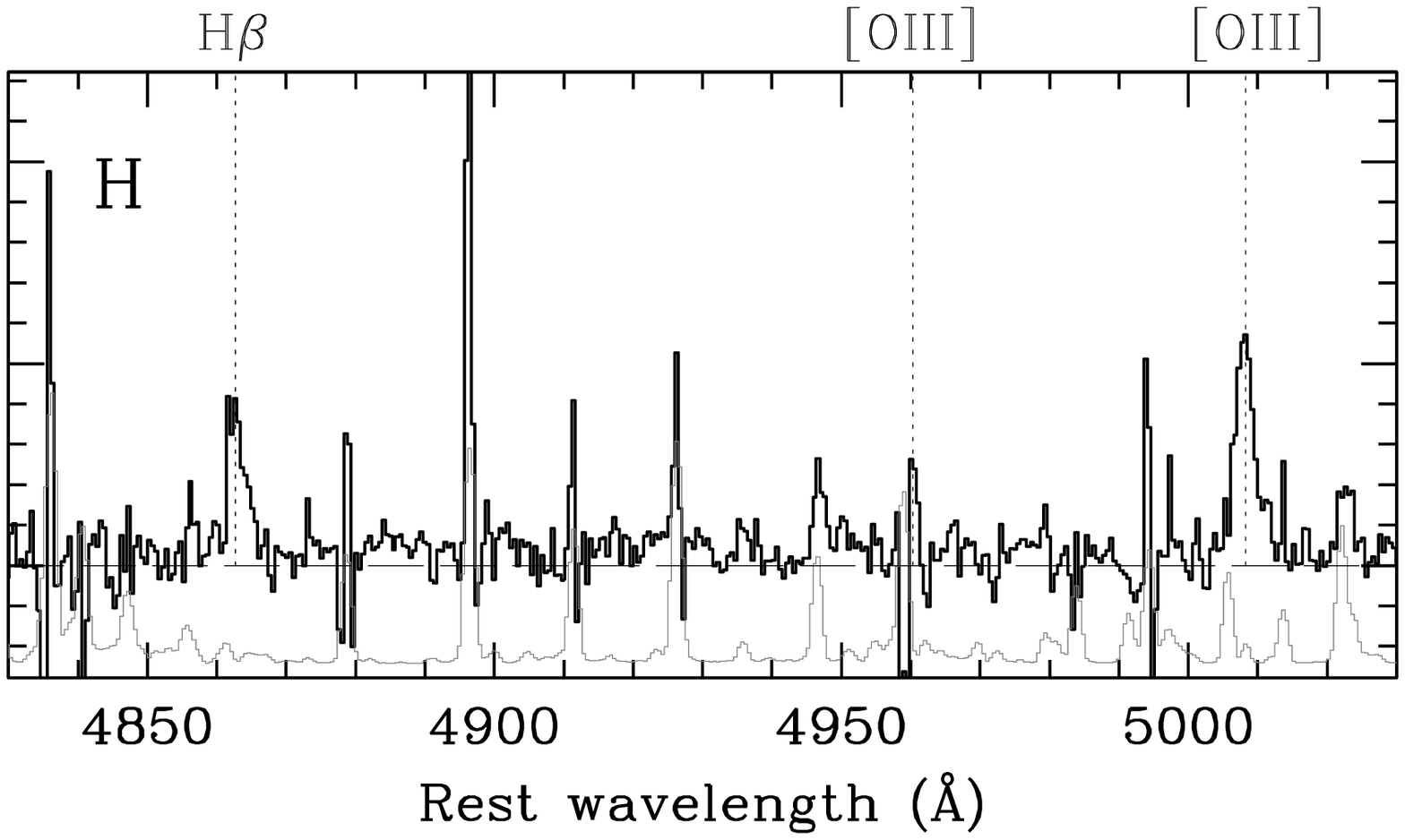}\includegraphics[height=1.35in]{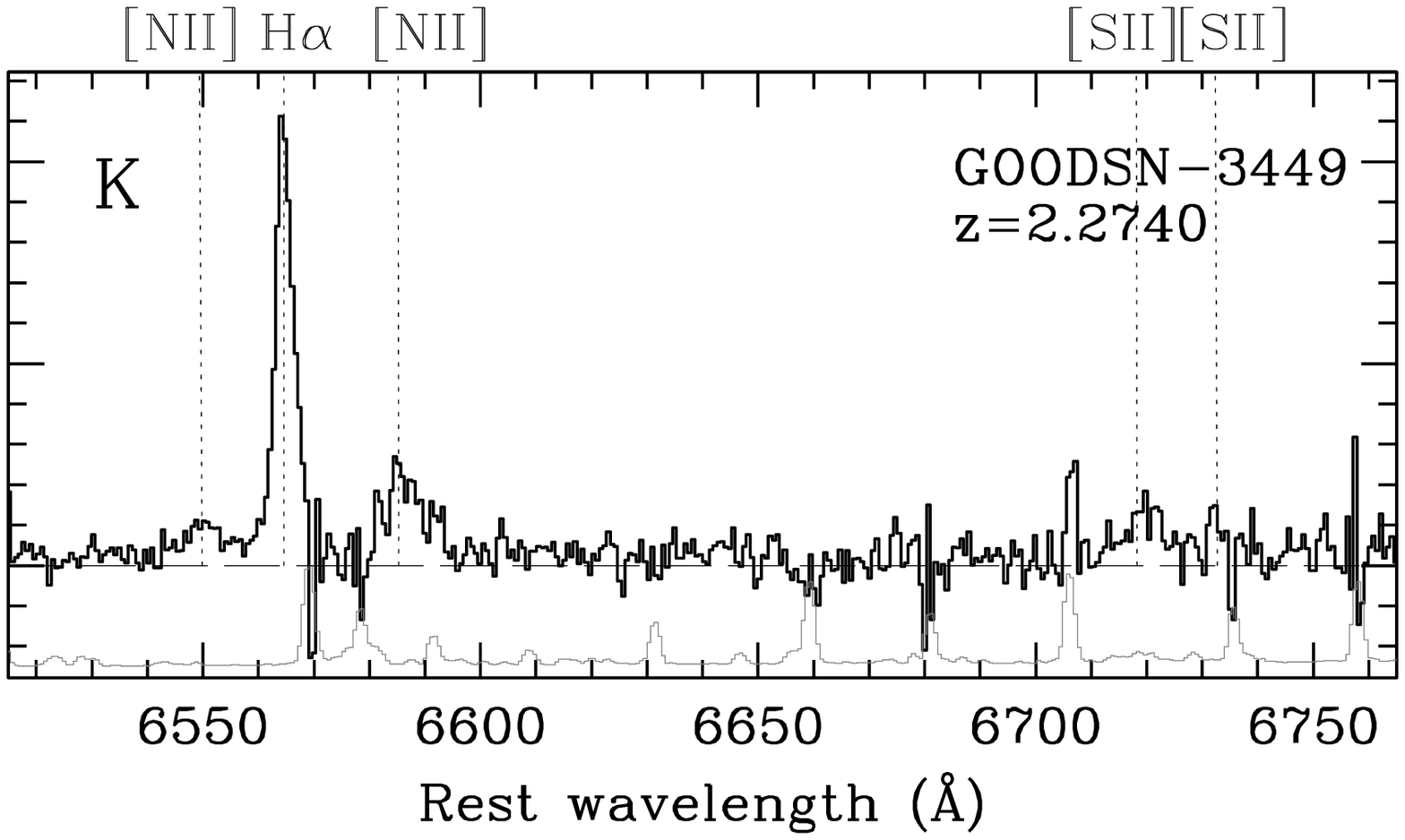}}
\caption{Example flux-calibrated J-band (left), H-band (center), and K-band (right) MOSFIRE spectra for two galaxies
from the MOSDEF sample, demonstrating the range of observed excitation properties. The top panels
show the emission-line spectrum of AEGIS-1361 ($z=2.1836$), a MOSDEF galaxy in the lowest
quartile of stellar mass ($M_*=10^{9.53} M_{\odot}$), with lower  [NII]$\lambda 6585$/H$\alpha$,
and higher [OIII]$\lambda5007$/H$\beta$ and $O_{32}$. The bottom panels show the emission-line spectrum of GOODSN-3449
($z=2.2740$), a MOSDEF galaxy in the highest
quartile of stellar mass ($M_*=10^{10.67} M_{\odot}$), with higher [NII]$\lambda 6585$/H$\alpha$,
and lower [OIII]$\lambda5007$/H$\beta$ and $O_{32}$. In each set of panels, the spectra from all filters
are on the same vertical scale, which is set by the height of the strongest emission line
([OIII]$\lambda5007$ for AEGIS-1361 and H$\alpha$ for GOODSN-3449). The spectra are unsmoothed and unbinned, with
the wavelength scale shifted to the rest frame. The $1\sigma$ error spectrum is plotted in grey, and offset vertically from
the corresponding science spectrum in each panel.}
\label{fig:plotspec}
\end{figure*}

Several key galaxy properties were derived from a combination of our MOSFIRE measurements
and existing ancillary data. Nebular extinction, $E(B-V)_{neb}$, was estimated
based on the stellar-absorption-corrected H$\alpha$/H$\beta$ ratio, assuming an intrinsic
ratio of 2.86 \citep[appropriate for $T_e = 10,000$~K;][]{osterbrock1989}
and using the dust-attenuation curve of \citet{cardelli1989}.
Star-formation rates (SFRs) were estimated from dust-corrected H$\alpha$ luminosities,
using the calibration of \citet{kennicutt1998} converted to a \citet{chabrier2003}
IMF. Stellar masses ($M_*$) were estimated using the pre-existing
multi-wavelength photometry assembled by the 3D-HST team \citep{skelton2014}, with spectroscopic
redshifts fixed by our MOSFIRE measurements. For SED fitting, we used the program FAST \citep{kriek2009},
assuming the stellar population synthesis models of \citet{conroy2009} and a \citet{chabrier2003}
IMF. Specific SFRs (sSFRs) were estimated as the ratio between the dust-corrected H$\alpha$
SFRs and stellar masses. Finally, as described in Freeman et al., in prep,
we estimated galaxy sizes by counting the number
of pixels above a fixed rest-frame-UV surface-brightness threshold in PSF-deconvolved {\it HST}/ACS images
of our targets. This estimate of size should more accurately reflect the area of active star formation
(either contiguous or clumpy)
than the effective radius derived from a \citet{sersic1968} fit to either rest-frame optical or UV profiles.
We then divided dust-corrected H$\alpha$ SFRs by rest-frame UV areas to estimate SFR surface
densities ($\Sigma_{SFR}$).

\subsection{Sample}
\label{sec:observations-sample}
In this paper, we explore the excitation properties of $z\sim 2.3$ star-forming
galaxies, as probed by their rest-frame optical emission-line ratios.
We consider the [OIII]$\lambda 5007$/H$\beta$ vs. [NII]$\lambda 6585$/H$\alpha$ and  [OIII]$\lambda5007$/H$\beta$
vs. [SII]/H$\alpha$ ``BPT" diagrams, 
as well as that of $O_{32}$ vs. $R_{23}$. 
In all plots of $O_{32}$ and $R_{23}$, we have estimated the sum of [OIII]$\lambda\lambda4959,5007$
fluxes as $(1+1/2.98)\times$[OIII]$\lambda5007$, given that the [OIII]$\lambda5007$/[OIII]$\lambda4959$ line ratio
is fixed \citep{storey2000} and [OIII]$\lambda5007$ typically offers a higher S/N 
measurement than [OIII]$\lambda4959$.

\begin{figure*}[t!]
\centerline{\includegraphics[height=3.5in]{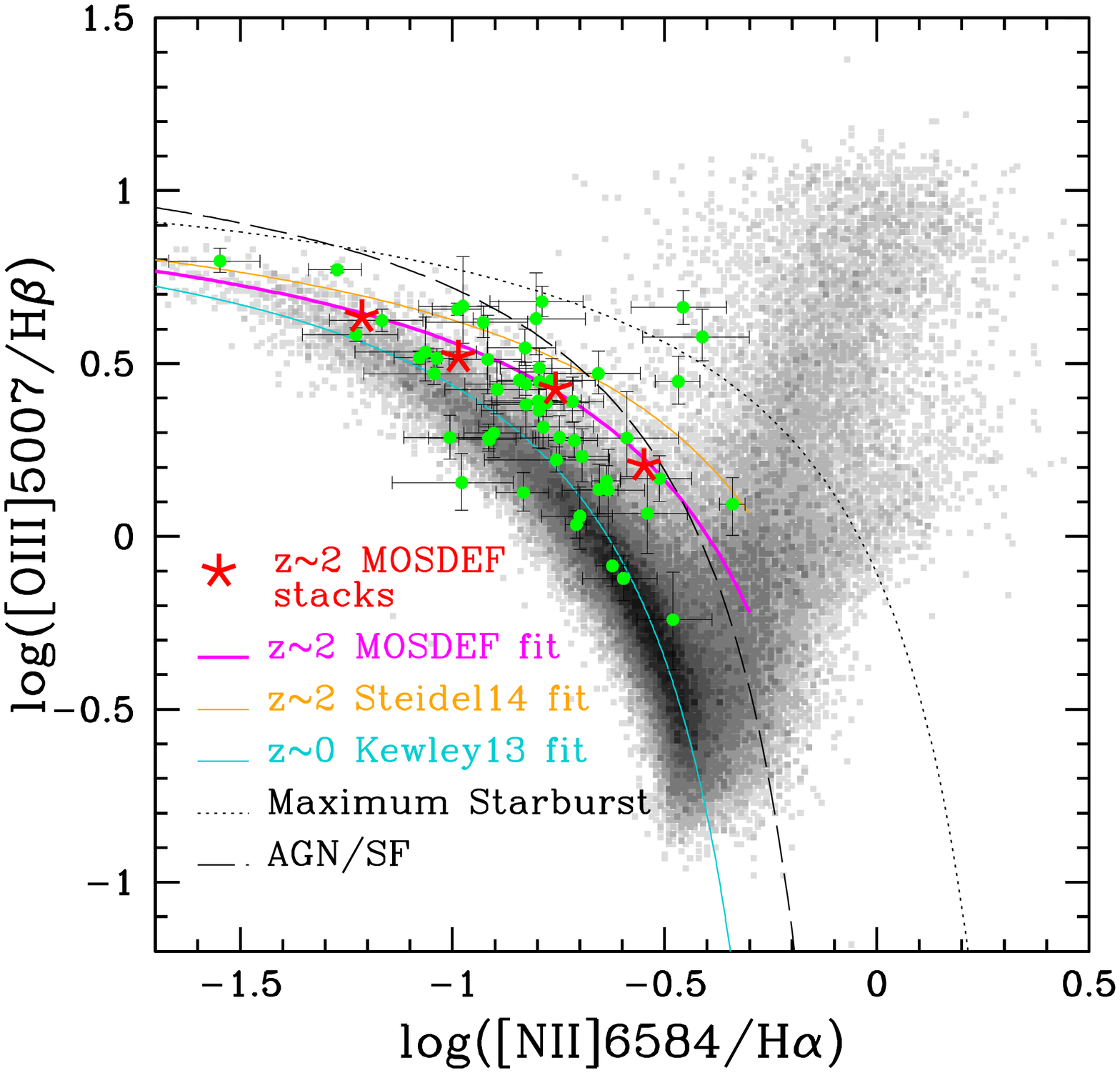}\includegraphics[height=3.5in]{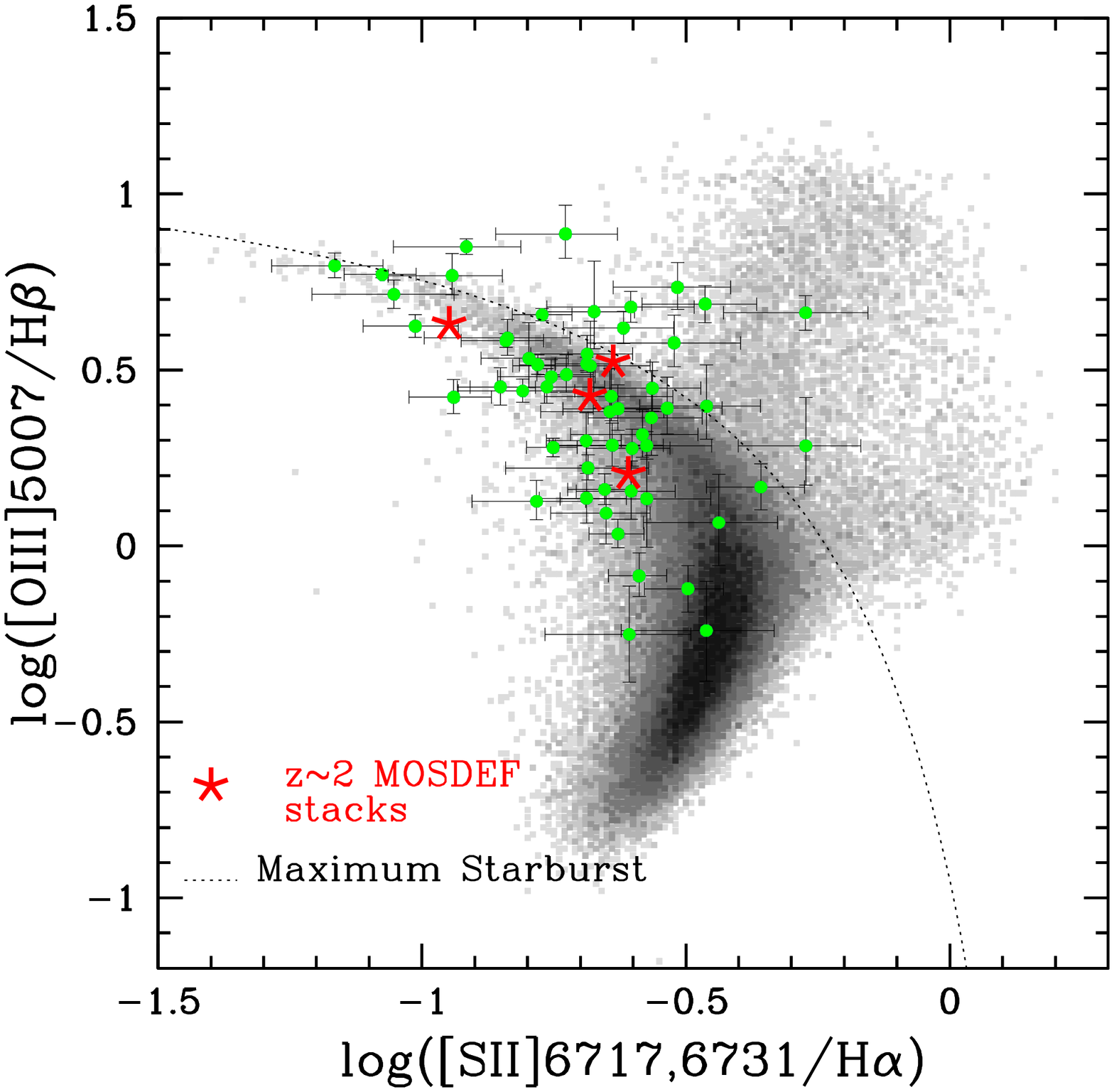}}
\caption{Left: [OIII]$\lambda 5007$/H$\beta$ vs. [NII]$\lambda 6585$/H$\alpha$ BPT diagram for $z\sim 2.3$ MOSDEF galaxies. Green points indicate
the sample of 53 MOSDEF galaxies with $\geq 3\sigma$ detections in H$\beta$,
[OIII]$\lambda5007$, H$\alpha$ and [NII]$\lambda6584$. The greyscale histogram
corresponds to the distribution of local SDSS galaxies. Large red stars represent
measurements of stacks binned by stellar mass of all MOSDEF galaxies with coverage of the relevant
emission lines regardless of whether or not lines were detected, indicating that plotting
MOSDEF detections only does not result in significant bias. Stacks of increasing mass
are characterized by lower [OIII]$\lambda5007$/H$\beta$ and higher [NII]$\lambda6584$/H$\alpha$. 
The black dotted curve is the ``maximum starburst" line from \citet{kewley2001}, while the black
dashed curve is an empirical AGN/star-formation threshold from \citet{kauffmann2003}. The cyan curve
indicates the $z\sim 0$ star-forming locus \citep{kewley2013}, while the magenta curve is the
best-fit to the $z\sim 2.3$ MOSDEF sequence. The orange curve is the best-fit
to $z\sim 2.3$ UV-selected galaxies from \citet{steidel2014}. The fit to the $z\sim 2.3$
MOSDEF sample is not as significantly offset from the $z\sim 0$ star-forming
sequence as is the fit describing the \cite{steidel2014} sample. Right:  [OIII]$\lambda5007$/H$\beta$
vs. [SII]/H$\alpha$ BPT diagram
for $z\sim 2.3$ MOSDEF galaxies. Symbols are as in the left-hand panel. In the  [OIII]$\lambda5007$/H$\beta$
vs. [SII]/H$\alpha$ diagram,
the $z\sim 2.3$ sample scatters symmetrically around the $z\sim 0$ star-forming sequence.
}
\label{fig:O3N2O3S2}
\end{figure*}

In order to plot MOSDEF galaxies in a given emission-line diagnostic space, we require spectral coverage
of all of the relevant emission lines. Furthermore, we restrict the sample to objects with
no evidence for AGN activity based on either X-ray luminosity or rest-frame near-IR colors
\citep{coil2014}. To allow for systematic differences in the emission-line
ratios of local and high-redshift galaxies, we do not additionally apply commonly adopted
rest-frame optical AGN criteria for sample selection \citep[e.g.,][]{kauffmann2003,kewley2001}. 
Out of a parent MOSDEF sample
of 133 non-AGN targets with robust MOSFIRE spectroscopic redshifts at $1.9 \leq z \leq 2.7$,\footnote{Although
the target redshift range is $2.09 \leq z \leq 2.61$, some targets with only photometric redshifts
prior to MOSFIRE observations turned out to be at slightly lower redshift when spectroscopically
confirmed. We furthermore included serendipitous objects falling on slits and roughly within the target
redshift range. We therefore use a larger redshift interval than initially designed to select galaxies
for this study.} 
118, 115, and 89 objects, respectively, have coverage of the lines
featured in the [OIII]$\lambda 5007$/H$\beta$ vs. [NII]$\lambda 6585$/H$\alpha$,  [OIII]$\lambda5007$/H$\beta$
vs. [SII]/H$\alpha$, and $O_{32}$ vs. $R_{23}$ plots.\footnote{The $O_{32}$ vs. $R_{23}$ sample is slightly smaller
than the [OIII]$\lambda 5007$/H$\beta$ vs. [NII]$\lambda 6585$/H$\alpha$ and  [OIII]$\lambda5007$/H$\beta$
vs. [SII]/H$\alpha$ samples, since two of the early MOSDEF masks observed did not
have J-band coverage of [OII].} As shown in Figure~\ref{fig:zhist_samples}, the resulting
redshift distributions are very similar for the three samples, with $\langle z \rangle = 2.3$.
Furthermore, since a restriction of our analysis to the 88 galaxies with coverage in all six strong
features yields qualitatively similar results, we opt to maximize the 
sample size for each parameter space. The galaxies in our $z\sim 2.3$ excitation analysis
span the range $10^{9.05}-10^{11.48} M_{\odot}$ in stellar mass, and $4-180 M_{\odot}\mbox{ yr}^{-1}$
in dust-corrected H$\alpha$ SFR. Figure~\ref{fig:plotspec} shows examples of MOSFIRE J, H, and K
spectra for two galaxies drawn from the MOSDEF sample, including AEGIS-1361 ($z=2.1836$), 
a galaxy in the lowest-mass quartile of the sample with $M_*=10^{9.53} M_{\odot}$, and GOODSN-3449 ($z=2.2740$),
a galaxy in the highest-mass quartile with $M_*=10^{10.67} M_{\odot}$. These spectra provide a typical representation of
how the pattern of rest-frame optical emission lines varies with stellar mass.

For the study of redshift evolution in each emission-line
diagnostic space, we construct $z\sim 0$ comparison samples from the Sloan Digital Sky 
Survey \citep[SDSS;][]{york2000} Data Release 7 (DR7) catalog \citep{abazajian2009}. Emission-line
measurements (corrected for underlying stellar absorption) and galaxy properties 
are drawn from the MPA-JHU catalog of measurements for SDSS DR7.\footnote{Available at 
http://www.mpa-garching.mpg.de/SDSS/DR7/.} Specifically, 
we select SDSS galaxies at $0.04 \leq z \leq 0.10$ to reduce aperture effects  (though they are still present, even at $z=0.10$), and require $5\sigma$ 
detections in the rest-frame optical emission lines featured in each diagnostic diagram.
For the [OIII]$\lambda 5007$/H$\beta$ vs. [NII]$\lambda 6585$/H$\alpha$ and  [OIII]$\lambda5007$/H$\beta$
vs. [SII]/H$\alpha$ diagnostic diagrams, we retain all SDSS galaxies, regardless of where
they fall relative to the AGN criterion of \citet{kauffmann2003}. For the $O_{32}$ vs. $R_{23}$ diagram, in which
we focus solely on the ionization parameter and metallicity for star-forming galaxies,
we reject local AGNs with the \citeauthor{kauffmann2003} criterion. SDSS comparison
samples in [OIII]$\lambda 5007$/H$\beta$ vs. [NII]$\lambda 6585$/H$\alpha$ and  [OIII]$\lambda5007$/H$\beta$
vs. [SII]/H$\alpha$ spaces contain $\sim 86,000$ galaxies, while the $O_{32}$ vs. $R_{23}$ comparison
sample contains $\sim 65,000$ galaxies. These comparison samples are well matched to the corresponding MOSDEF
samples in terms of median stellar mass, yet they are characterized by median SFRs that are 
a factor of $\sim 20$ lower.

\section{The Excitation Properties of $z\sim 2.3$ Galaxies}
\label{sec:results}
\subsection{Diagnostic Diagrams}
\label{sec:results-diagnostics}

\begin{figure*}[t!]
\centerline{\includegraphics[height=3.5in]{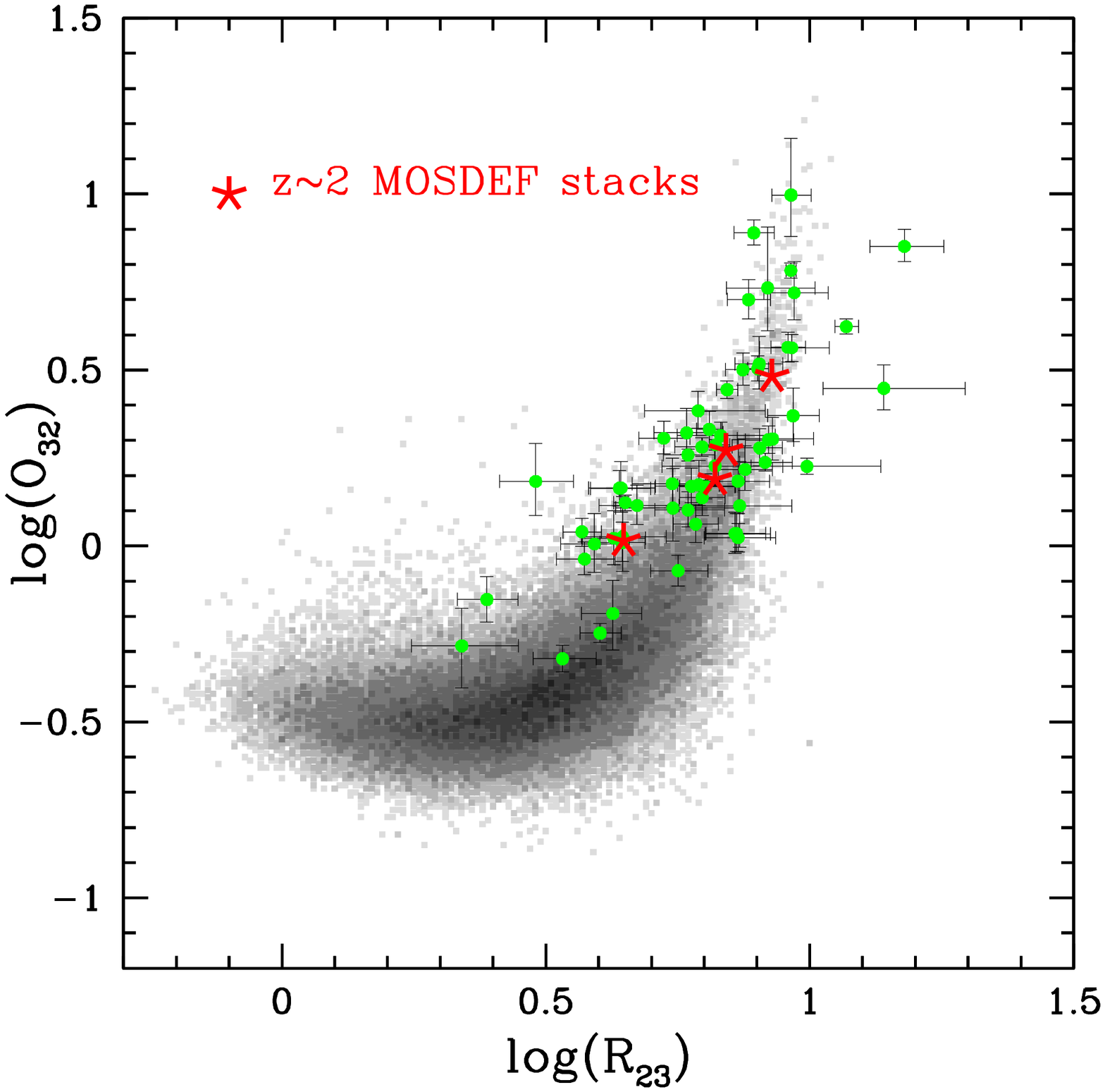}\includegraphics[height=3.5in]{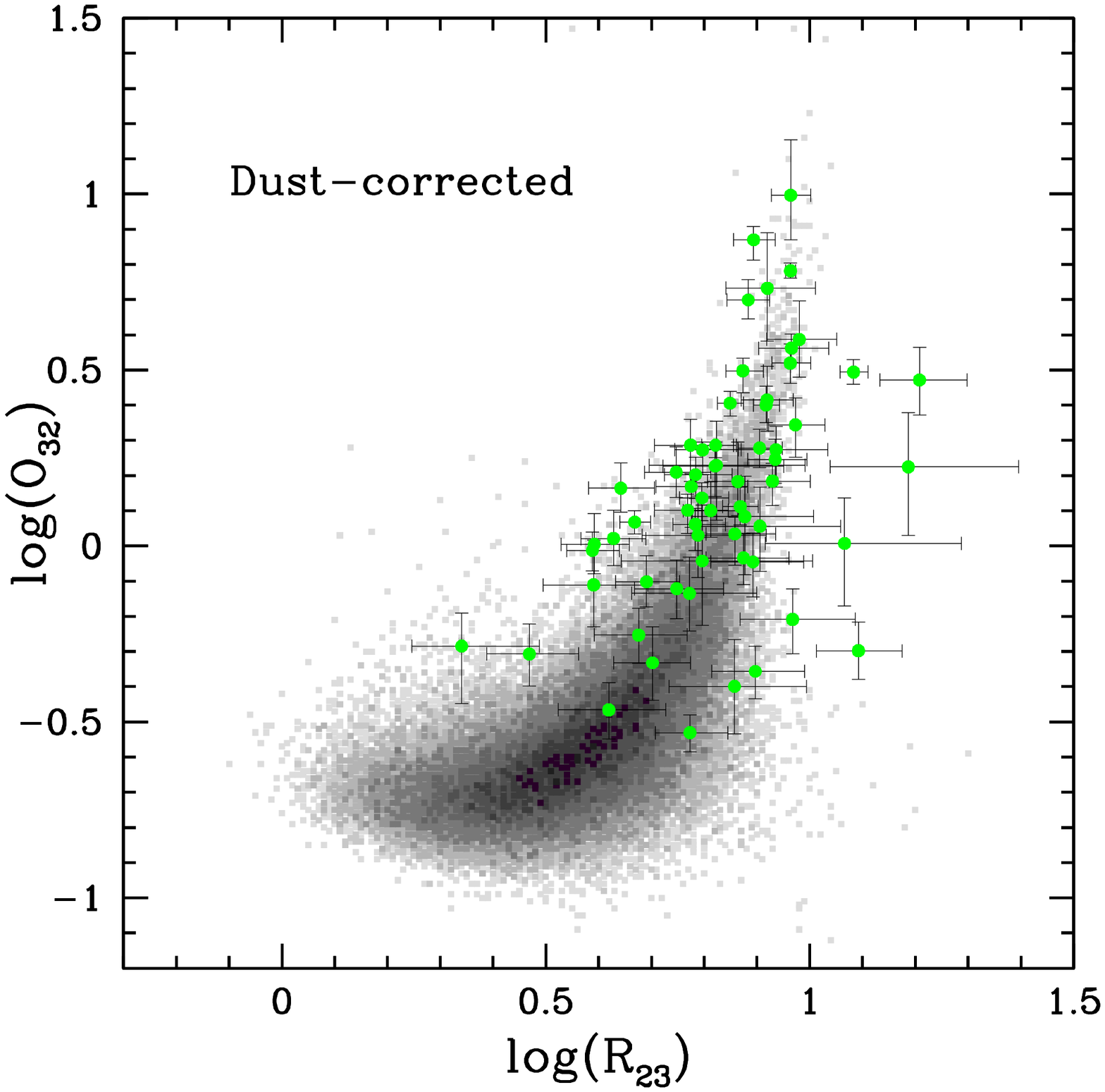}}
\caption{Left: $O_{32}$ vs. $R_{23}$ excitation diagram for $z\sim 2.3$ MOSDEF galaxies. Symbols
are as in Figure~\ref{fig:O3N2O3S2} for individual MOSDEF points and stacks in stellar mass,
as well as SDSS galaxies. Stacks of increasing stellar mass are characterized by lower $O_{32}$
values. The $z\sim 2.3$ MOSDEF sample closely follows the high-$O_{32}$,
high $R_{23}$ portion of the local excitation sequence, without evidence for a systematic
offset in this parameter space.  Right: $O_{32}$ vs. $R_{23}$ diagram, corrected for dust. Dust corrections
are applied to both $z\sim 2.3$ MOSDEF and SDSS galaxies based on the observed H$\alpha$/H$\beta$ ratio
(corrected for underlying stellar absorption) and assuming the \citet{cardelli1989} attenuation
curve. Dust corrections shift the observed distribution towards higher $R_{23}$ and lower $O_{32}$
values, but our conclusion that the $z\sim 2.3$ MOSDEF sample follows the high-excitation portion
of the local sample does not change.}
\label{fig:O32R23}
\end{figure*}

The large, rest-frame-optically-selected sample of $z\sim 2.3$ MOSDEF galaxies with coverage of multiple
strong rest-frame optical emission lines allows for an unprecedented investigation of the excitation
properties in high-redshift star-forming regions. Crucial to understanding the underlying
excitation mechanisms is the consideration of multiple different diagnostic diagrams.
Here we investigate the locations of $z\sim 2.3$ star-forming galaxies in the
[OIII]$\lambda 5007$/H$\beta$ vs. [NII]$\lambda 6585$/H$\alpha$,  [OIII]$\lambda5007$/H$\beta$
vs. [SII]/H$\alpha$, and $O_{32}$ vs. $R_{23}$ emission-line diagnostic diagrams.

In Figure~\ref{fig:O3N2O3S2},
we show the MOSDEF $z\sim 2.3$ galaxies in the space of [OIII]$\lambda 5007$/H$\beta$ vs. [NII]$\lambda 6585$/H$\alpha$ (left) and  [OIII]$\lambda5007$/H$\beta$
vs. [SII]/H$\alpha$ (right).
We only plot galaxies with 3$\sigma$ or greater detections in the relevant emission line fluxes,
leading to a sample of 53 galaxies in the [OIII]$\lambda 5007$/H$\beta$ vs. [NII]$\lambda 6585$/H$\alpha$ diagram and 56 in the space of  [OIII]$\lambda5007$/H$\beta$ vs. [SII]/H$\alpha$.
Given the small number (4) of objects with [OIII]$\lambda 5007$/H$\beta$ upper limits in both of these samples, 
the requirement of [OIII]$\lambda 5007$ detections does not result in a significant bias against galaxies with 
low [OIII]$\lambda 5007$/H$\beta$ ratios typical of metal-rich star-forming regions. We return to the question
of sample selection bias at the end of this section.
For comparison, we plot SDSS galaxies in the same diagrams, along with a theoretical prediction
of the $z\sim 0$ [OIII]$\lambda 5007$/H$\beta$ vs. [NII]$\lambda 6585$/H$\alpha$ star-forming sequence from \citet{kewley2013}, which provides a good
fit to the local data.

As initially suggested by the small high-redshift samples featured 
in \citet{shapley2005b,erb2006a,liu2008}, and recently 
confirmed by \citet{steidel2014} using a much larger sample, the set of $z\sim 2.3$
MOSDEF points is clearly offset on average from the $z\sim 0$ star-forming sequence
in the [OIII]$\lambda 5007$/H$\beta$ vs. [NII]$\lambda 6585$/H$\alpha$ diagram.\footnote{We note that there are two galaxies in the MOSDEF
[OIII]$\lambda 5007$/H$\beta$ vs. [NII]$\lambda 6585$/H$\alpha$ sample that lie above the \citet{kewley2001} ``maximum Starburst" curve
in a region populated primarily by AGNs in the local universe. There is no evidence
based on X-ray or rest-frame near-IR properties that these objects are AGNs \citep{coil2014},
therefore we retain them in our sample. We note that the functional form
derived for the excitation sequence of the $z\sim 2.3$ MOSDEF sample is insensitve
to the inclusion or exclusion of these two objects.}
With the adoption of the same functional form as that
used by \citet{kewley2013} and \citet{steidel2014}, the best-fit to the $z\sim 2.3$ MOSDEF sample
star-forming sequence is:

\begin{equation}
\log(\mbox{[OIII]}/\mbox{H}\beta) = \frac{0.67}{\log(\mbox{[NII]}/\mbox{H}\alpha)-0.20} + 1.12
\label{eq:o3n2}
\end{equation}

Compared with the fit to the $z\sim 2.3$ UV-selected sample of \citet{steidel2014} (orange
curve in Figure~\ref{fig:O3N2O3S2}), or the locus of emission-line-selected 
galaxies from \citet{masters2014} at $z\sim 1.5$ and $2.2$,
the sequence of rest-frame-optically selected MOSDEF galaxies 
as a whole is not as offset from the local one. This difference underscores
the importance of sample selection at high-redshift in determining the region occupied in the [OIII]$\lambda 5007$/H$\beta$ vs. [NII]$\lambda 6585$/H$\alpha$ 
diagram. Although the \citet{steidel2014} sample has a similiar median stellar mass to that
of the $z\sim 2.3$ MOSDEF sample ($\sim 10^{10} M_{\odot}$), it spans a wider range of stellar masses
(from $10^{8.6} M_{\odot}$ to $10^{11.4}$) than the MOSDEF sample. Furthermore, at the lowest masses ($<10^{9.5} M_{\odot}$),
galaxies in the \citeauthor{steidel2014} sample have higher average SFR and sSFR. The \citeauthor{masters2014}
sample is significantly lower in mass than the MOSDEF sample, with a median of $\sim 10^9 M_{\odot}$.
Given the similar range of SFRs in the \citeauthor{masters2014} and MOSDEF samples, the \citeauthor{masters2014}
sample is characterized by significantly higher sSFR on average.
It is crucial to understand the differences in galaxy physical properties probed
by the various $z\sim 2$ samples, and how they translate into differences in the
[OIII]$\lambda 5007$/H$\beta$ vs. [NII]$\lambda 6585$/H$\alpha$ diagram. It is
these intrinsic physical properties that modulate the integrated line ratios for 
a galaxy, as opposed to simply the redshift at which a galaxy is observed.
Along these lines, we investigate possible
causes of the difference in [OIII]$\lambda 5007$/H$\beta$ vs. [NII]$\lambda 6585$/H$\alpha$ star-forming sequences in 
section~\ref{sec:results-galprops}, with the separation of our 
sample according to various galaxy properties.

\begin{figure*}[t!]
\centerline{\includegraphics[height=2.4in]{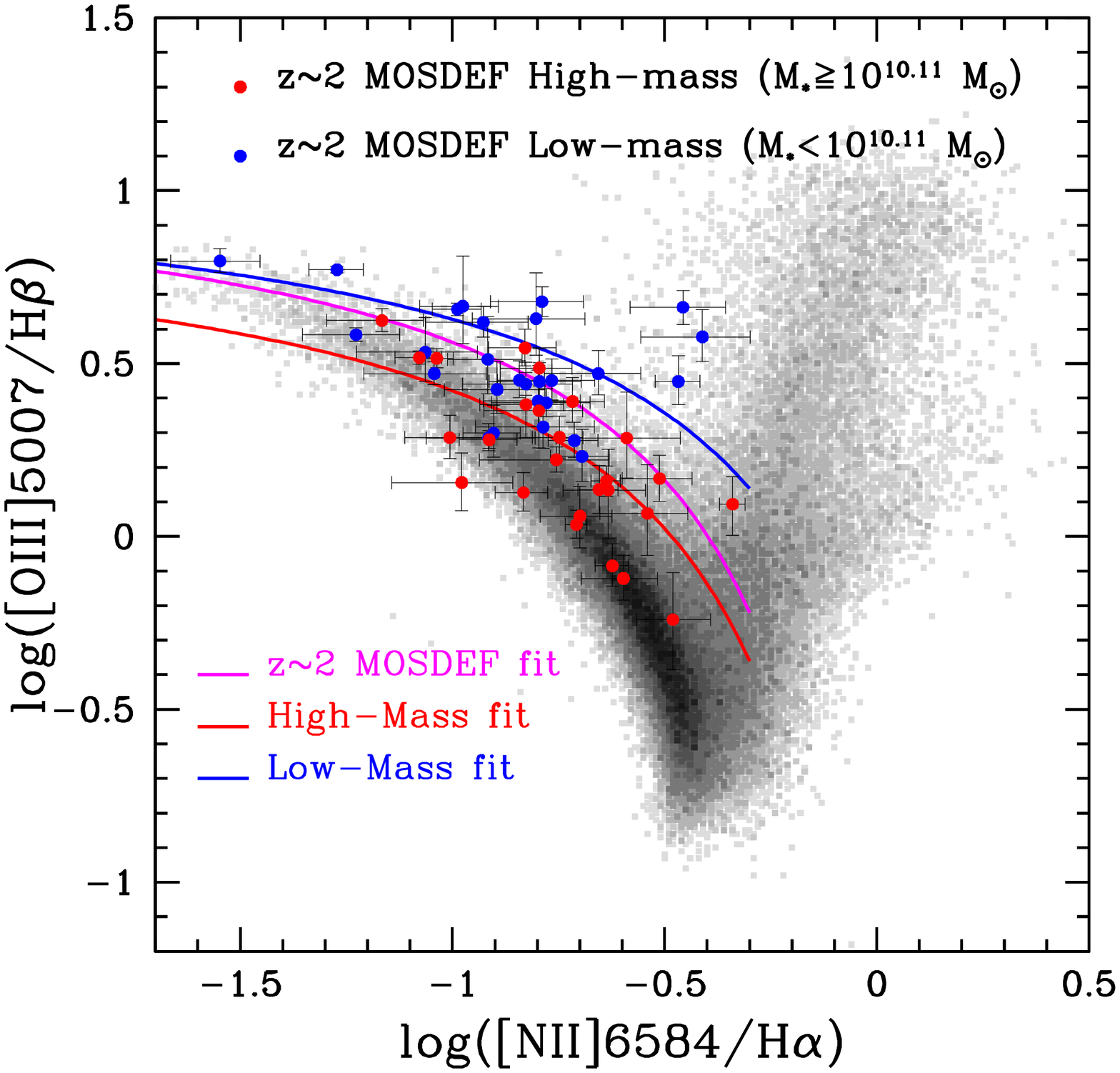}\includegraphics[height=2.4in]{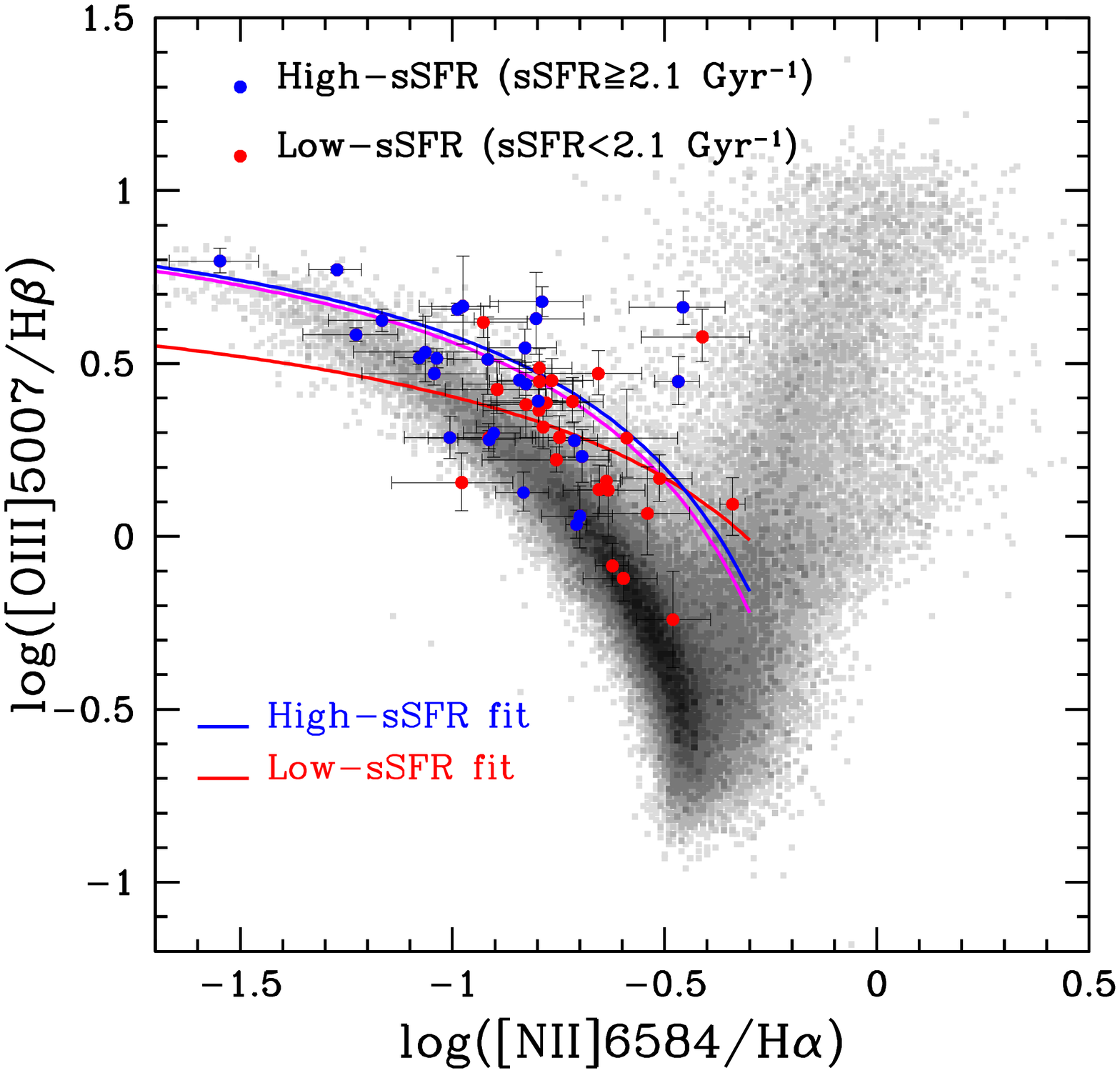}\includegraphics[height=2.4in]{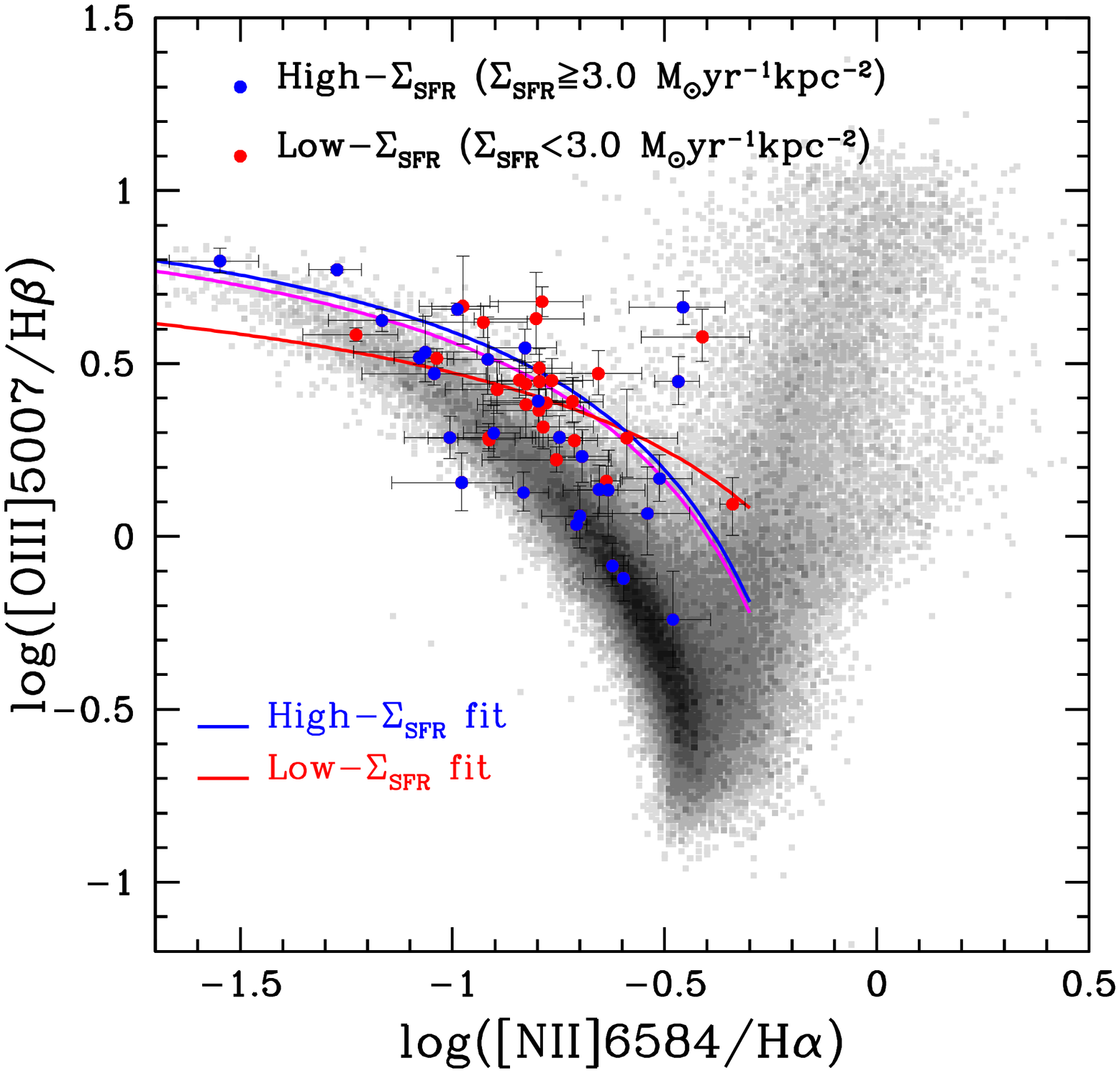}}
\caption{[OIII]$\lambda 5007$/H$\beta$ vs. [NII]$\lambda 6585$/H$\alpha$ BPT diagram, color-coded by galaxy property. In each panel blue and red points
correspond to the $z\sim 2.3$ MOSDEF sample, while the grey histogram indicates the $z\sim 0$
SDSS sample. A fit to the full $z\sim 2.3$ MOSDEF sample is indicated in each panel with the magenta
curve, while fits to different subsamples are shown with red and blue curves.
Left: Separation by stellar mass. Blue [red] symbols represent
$z\sim 2.3$ MOSDEF galaxies with stellar masses below [above] the sample median. High-mass galaxies
follow the $z\sim 0$ star-forming sequence without significant evidence for an offset, while
low-mass galaxies are offset in either [OIII]/H$\beta$ or [NII]/H$\alpha$.
Center: Separation by sSFR. Blue [red] symbols represent
$z\sim 2.3$ MOSDEF galaxies with sSFR above [below] the sample median. The segregation
by sSFR at $z\sim 2.3$ is less clearly perpendicular to the $z\sim 0$ star-forming sequence
over the range populated by both ``high" and ``low" samples. Overall, the separation is more so
along the sequence, with low-sSFR galaxies shifted towards higher [NII]/H$\alpha$ values than
high-sSFR galaxies. Right: Separation by $\Sigma_{SFR}$. Blue [red] symbols represent
$z\sim 2.3$ MOSDEF galaxies with $\Sigma_{SFR}$ above [below] the sample median. The $z\sim 2.3$ MOSDEF galaxies
are very well-mixed as a function of $\Sigma_{SFR}$, with no obvious segregation.}
\label{fig:O3N2props}
\end{figure*}

The [OIII]$\lambda5007$/H$\beta$ vs. [SII]/H$\alpha$ BPT diagram has also been used to discriminate between
star-forming galaxies and AGNs, and star-forming galaxies form a fairly
well-defined sequence in this space as well. \citet{masters2014} constructed
a composite rest-frame optical spectrum from 24 emission-line galaxies at
$\langle z\rangle=1.85$, finding no systematic offset in the [SII]/H$\alpha$ vs.
[OIII]$\lambda5007$/H$\beta$ ratio relative to the metal-poor tail
of the local star-forming sequence \citep[see also][]{dominguez2013}. Based on our larger sample of individual
MOSDEF measurements, we confirm the lack of a systematic offset between the 
high-redshift galaxies and the local sequence in the [OIII]$\lambda5007$/H$\beta$ vs. [SII]/H$\alpha$ diagram. Indeed, the high-redshift data
points scatter roughly symmetrically around the star-forming SDSS points.

In addition to the two BPT diagrams, we consider the space of $O_{32}$ vs. $R_{23}$,
shown in  Figure~\ref{fig:O32R23} both with (right) and without (left) nebular extinction
corrections applied. The \citet{cardelli1989} attenuation law was used to correct emission-line 
fluxes based on the observed Balmer decrement. This attenuation curve is appropriate for correcting
nebular emission lines in the local universe. The attenuation law for nebular emission lines at high redshift
has yet to be established, though it's worth noting that we obtain the same key results if
the attenuation curve of \citet{calzetti2000} is used instead.
$O_{32}$ vs. $R_{23}$ diagrams have typically been used to infer simultaneously the metallicity
and ionization parameter of star-forming galaxies \citep[e.g.,][]{lilly2003,nakajima2013}, 
as both $O_{32}$ and $R_{23}$ are sensitive to a combination of these physical quantities.
In both panels, the SDSS sample is plotted along with the $z\sim 2.3$ MOSDEF sample.
As noted by \citet{nakajima2013} and \citet{hainline2009}, the sample of $z\geq 2$ galaxies with 
$O_{32}$ and $R_{23}$ measurements tends
to exhibit systematically higher $O_{32}$ values than the bulk of SDSS galaxies,
implying higher ionization parameters on average. The $z\sim 2.3$ MOSDEF
galaxies follow this same trend, occupying the high-$O_{32}$, high-$R_{23}$ portion
of the SDSS distribution. Yet, it is worth pointing out that there is no systematic {\it offset}
between the high-redshift and local points. The $z\sim 2.3$ MOSDEF sample simply occupies the
low-metallicity, high-ionization-parameter tail of the local distribution. 
\citet{nakajima2013} find that the sample of low-metallicity galaxies in SDSS
with direct oxygen abundance estimates from \citet{nagao2006} populate this same region
of $O_{32}$ vs. $R_{23}$ parameter space.

In summary, although the $z\sim 2.3$ MOSDEF sample is offset from the local star-forming
sequence in the space of [OIII]$\lambda 5007$/H$\beta$ vs. [NII]$\lambda 6585$/H$\alpha$, we find no evidence for a systematic offset in either
of the two other diagnostic diagrams considered -- i.e., [OIII]$\lambda5007$/H$\beta$ vs. [SII]/H$\alpha$ or $O_{32}$ vs. $R_{23}$. A few
notable improvements in our analysis over past work include the fact that we feature
a statistical sample of individual high-redshift [SII]/H$\alpha$ measurements in the [OIII]$\lambda5007$/H$\beta$ vs. [SII]/H$\alpha$ diagram,
correct Balmer emission lines for underlying stellar absorption, 
and apply nebular extinction corrections based on the Balmer decrement to line ratios including lines that
are significantly separated in wavelength (e.g., $O_{32}$ and $R_{23}$). 

\begin{figure*}[t!]
\centerline{\includegraphics[height=2.4in]{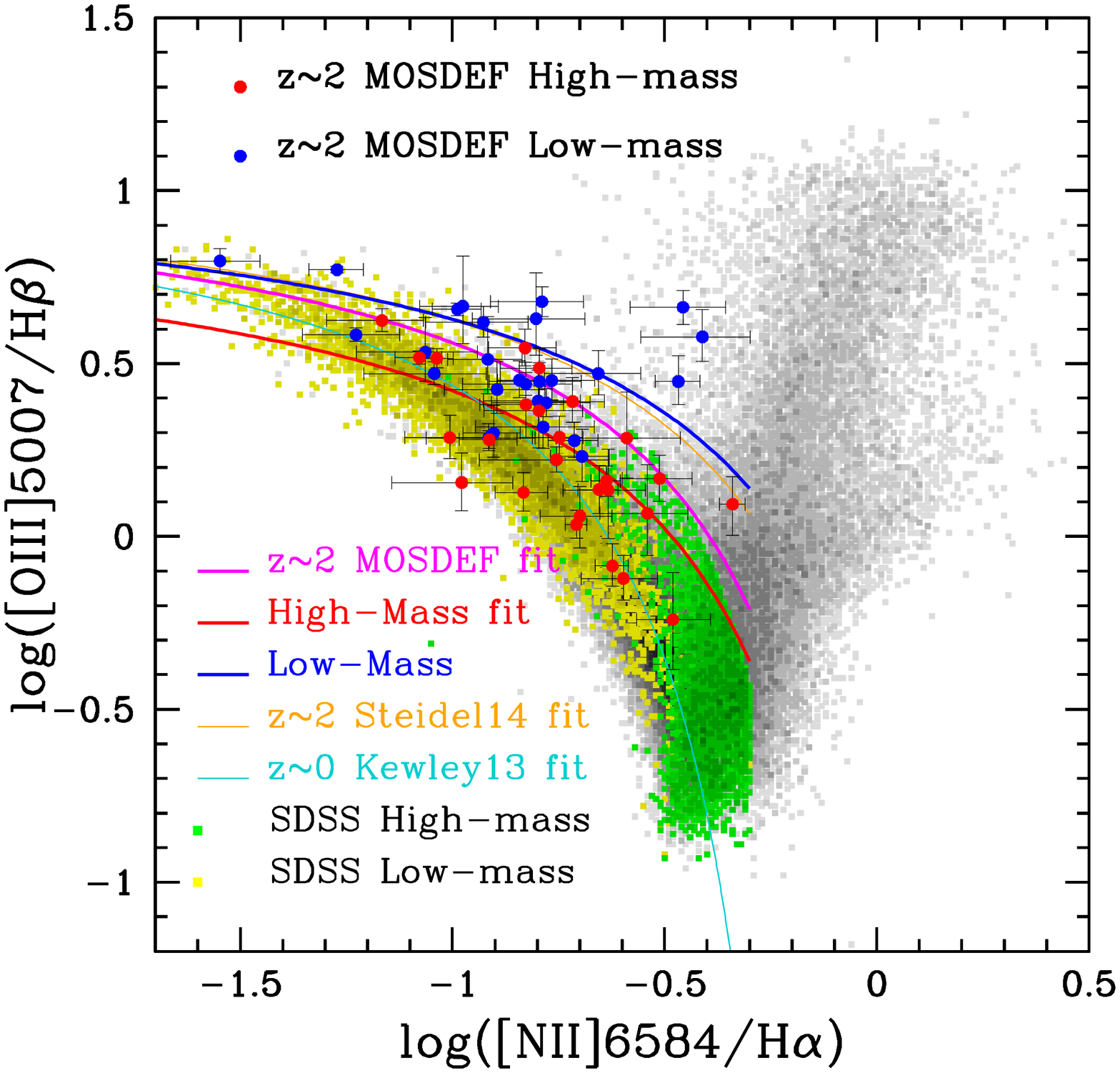}\includegraphics[height=2.4in]{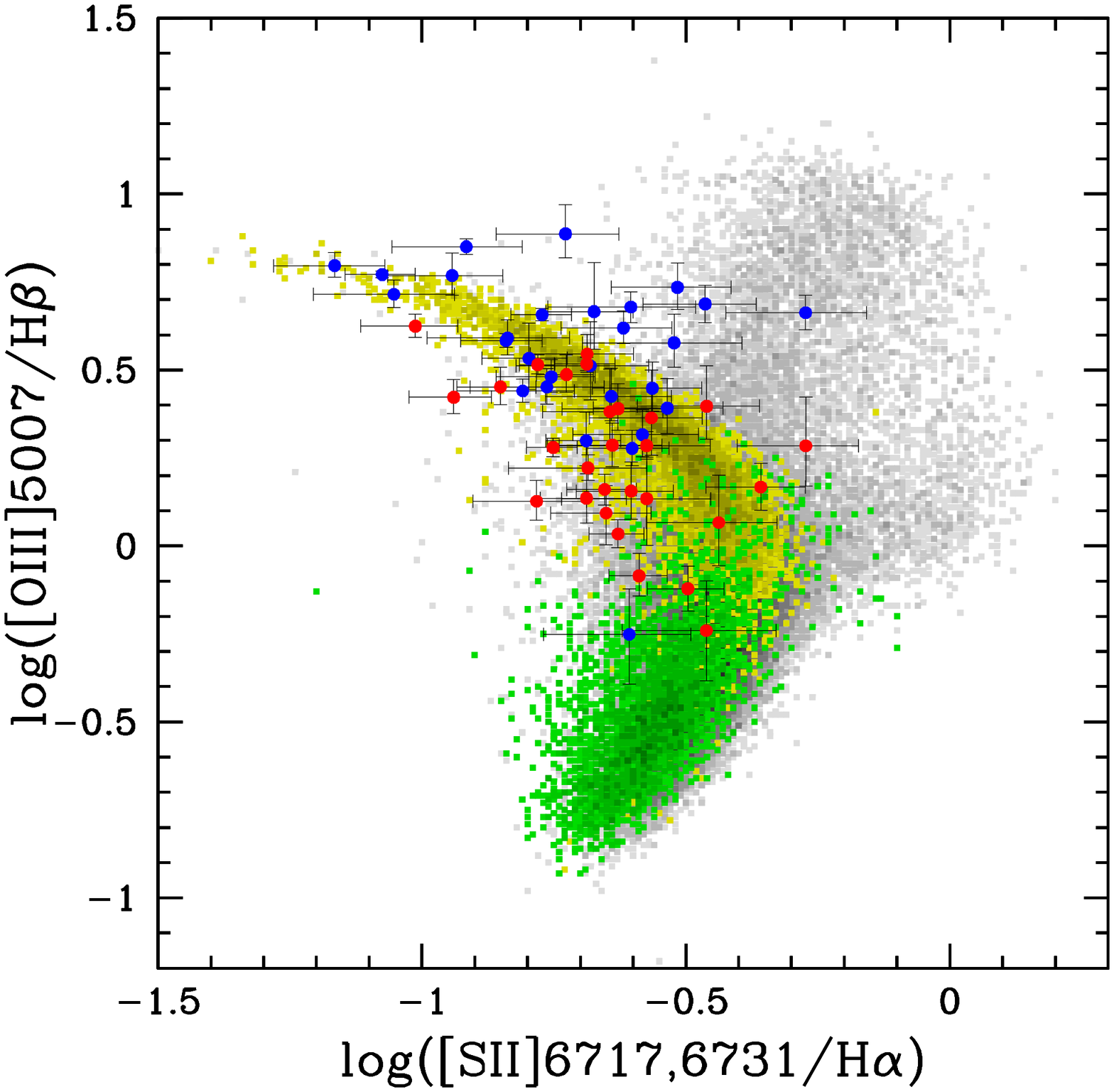}\includegraphics[height=2.4in]{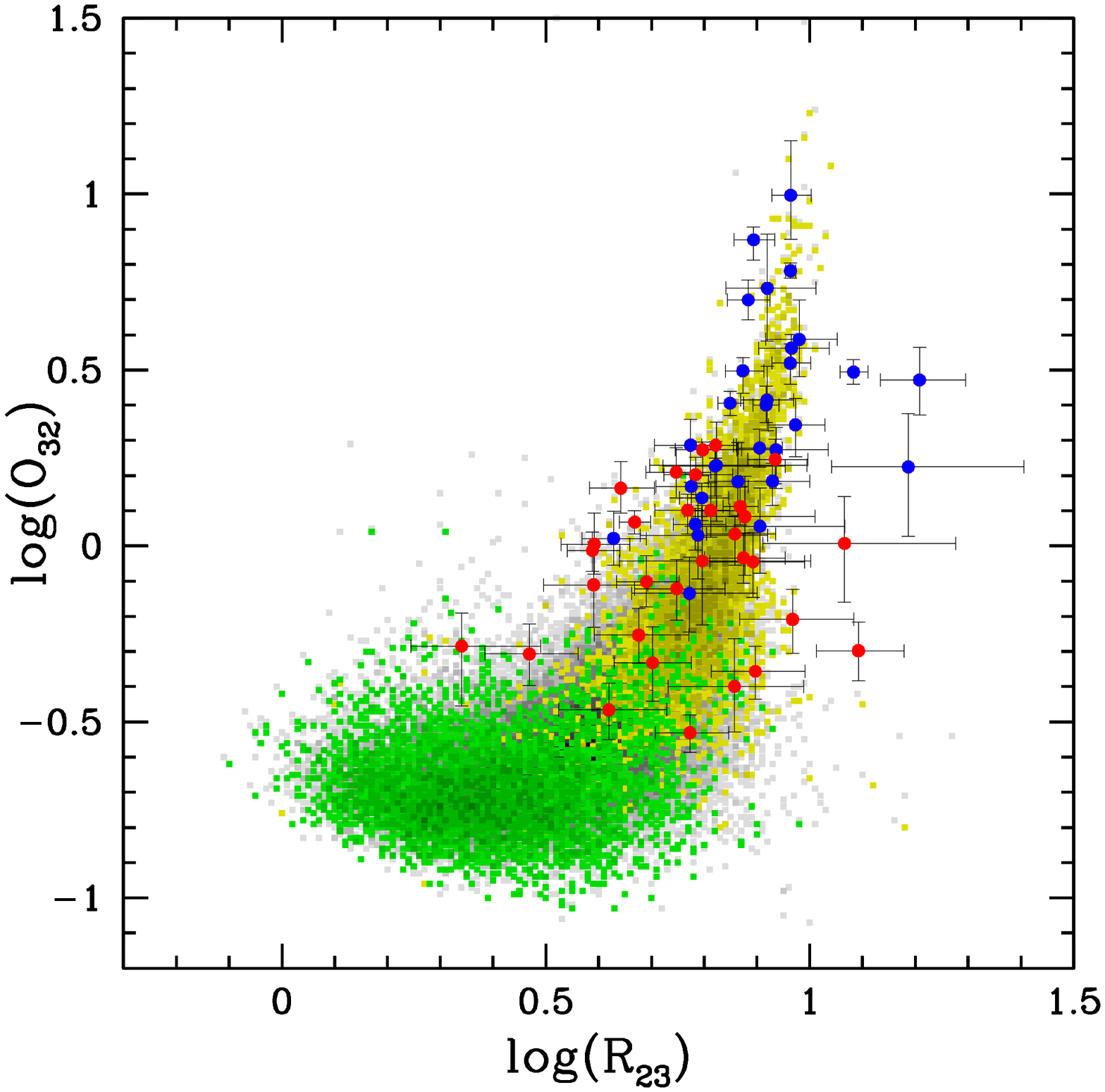}}
\caption{[OIII]$\lambda 5007$/H$\beta$ vs. [NII]$\lambda 6585$/H$\alpha$, [OIII]$\lambda5007$/H$\beta$ vs. [SII]/H$\alpha$, and $O_{32}$ vs. $R_{23}$ diagrams, segregated by stellar mass. In each panel,
$z\sim 2.3$ MOSDEF galaxies are indicated with blue (low-mass) and red (high-mass) points.
The median (dividing) mass is $10^{10.1} M_{\odot}$ for the [OIII]$\lambda 5007$/H$\beta$ vs. [NII]$\lambda 6585$/H$\alpha$ and [OIII]$\lambda5007$/H$\beta$ vs. [SII]/H$\alpha$ samples, and $10^{9.9} M_{\odot}$
for the $O_{32}$ vs. $R_{23}$ sample, with the difference for $O_{32}$ vs. $R_{23}$ resulting from the slightly different sample
emographics for this diagram (see Section~\ref{sec:observations-sample}).
SDSS star-forming galaxies \citep[i.e., those that fall below the curve of][]{kauffmann2003}
are shown with small yellow (low-mass, i.e., $M_*<10^{9.25} M_{\odot}$) and green (high-mass, i.e.,
$M_*>10^{10.52} M_{\odot}$) points. In each excitation diagram, low- and high-mass SDSS
galaxies separate along the excitation sequence. Left: [OIII]$\lambda 5007$/H$\beta$ vs. [NII]$\lambda 6585$/H$\alpha$ diagram. The $z\sim 0$ star-forming sequence,
as well as the fits to the overall $z\sim 2.3$ MOSDEF and \citet{steidel2014} samples are indicated
as in Figure~\ref{fig:O3N2O3S2}. Additionally, we show fits to the low-mass and high-mass
$z\sim 2.3$ subsamples using, respectively, blue and red curves. The fit to the low-mass portion
of the $z\sim 2.3$ MOSDEF sample is indistinguishable from the \citet{steidel2014} sequence,
while the fit to the high-mass $z\sim 2.3$ MOSDEF galaxies is very similar to the $z\sim 0$
star-forming sequence. Center: [OIII]$\lambda5007$/H$\beta$ vs. [SII]/H$\alpha$ diagram. Low-mass $z\sim 2.3$ MOSDEF galaxies
are offset towards higher [OIII]/H$\beta$ ratios, but both the low- and high-mass
portions of the MOSDEF sample overlap significantly with the $z\sim 0$ star-forming
sequence. Right: Dust-corrected $O_{32}$ vs. $R_{23}$ diagram. The low- and high-mass $z\sim 2.3$ MOSDEF
subsamples separate cleanly along the $O_{32}$ vs. $R_{23}$ excitation sequence, just
as the SDSS galaxies do.
}
\label{fig:alldiagnostics-mass}
\end{figure*}

One possible point of concern with our results stems from the fact that, thus far, we have
used only galaxies with detections in all of the relevant emission features in order
to trace out the locations of $z\sim 2.3$ galaxies in each diagnostic diagram.
This restriction leads to detections-only samples of 53, 56, and 61 galaxies,
respectively, in the [OIII]$\lambda 5007$/H$\beta$ vs. [NII]$\lambda 6585$/H$\alpha$, [OIII]$\lambda5007$/H$\beta$ vs. [SII]/H$\alpha$, and $O_{32}$ vs. $R_{23}$ diagrams. For the [OIII]$\lambda 5007$/H$\beta$ vs. [NII]$\lambda 6585$/H$\alpha$ and [OIII]$\lambda5007$/H$\beta$ vs. [SII]/H$\alpha$ diagrams,
the detections-only samples comprise 45-50\% of the MOSDEF targets with coverage of
these features (largely due to the difficulty of detecting faint [NII] and [SII]
lines), while this fraction is larger ($\sim70$\%) for $O_{32}$ vs. $R_{23}$. In order to assess
the potential bias of excluding galaxies with upper limits in line fluxes, we constructed
composite spectra for the full sample in each emission-line space in four bins of increasing stellar mass, 
with roughly equal numbers of galaxies in each bin. With stellar mass estimates for all MOSDEF galaxies, 
this property provides a natural basis for dividing the sample. 
For the [OIII]$\lambda 5007$/H$\beta$ vs. [NII]$\lambda 6585$/H$\alpha$ and [OIII]$\lambda5007$/H$\beta$ vs. [SII]/H$\alpha$ samples, we made composite
K and H-band spectra, covering, respectively H$\alpha$+[NII]$\lambda 6584$+[SII]$\lambda,\lambda 6717, 6731$,
and [OIII]$\lambda 5007$+H$\beta$, while, for $O_{32}$ vs. $R_{23}$, we made stacks in H and J,
covering, respectively, [OIII]$\lambda 5007$+H$\beta$ and [OII]$\lambda3727$.
Individual spectra in each filter were shifted to the rest frame, converted
from flux- to luminosity-density units (i.e., $\mbox{erg s}^{-1}\mbox{\AA}^{-1}$),
and normalized by H$\alpha$ luminosity in order to obtain the average line ratios
in each bin in stellar mass. Requiring the detection of H$\alpha$ to normalize
the individual spectra slightly reduces the 
[OIII]$\lambda 5007$/H$\beta$ vs. [NII]$\lambda 6585$/H$\alpha$, [OIII]$\lambda5007$/H$\beta$ vs. [SII]/H$\alpha$, and $O_{32}$ vs. $R_{23}$ sample sizes to 106, 105, and 84 galaxies, respectively,
($\geq 90$\% of the full samples with coverage of the relevant emission lines)
but should not result in a significant bias.
Stacked spectra for each stellar-mass
bin (in each filter) were then constructed as straight averages 
of the individual rest-frame, H$\alpha$-normalized spectra.
Emission-line ratios in each composite spectrum were measured by fitting Gaussian profiles to the H$\alpha$,
[NII], [OIII], and H$\beta$ features, and performing simple integration over [OII] and [SII].
Composite line ratios including H$\alpha$ and H$\beta$ were corrected for underlying Balmer absorption using
the median Balmer absorption correction among individually-detected galaxies in each bin.
For $O_{32}$ vs. $R_{23}$, we construct the stacks without correction for dust extinction given that we do 
not have individual H$\alpha$ and H$\beta$ detections for all objects being combined. However, these stacks should
still indicate how representative the observed $O_{32}$ and $R_{23}$ ratios are for individual
detections-only objects. As shown in Figures~\ref{fig:O3N2O3S2} and \ref{fig:O32R23}, the stacked
data points fall well within the distribution of individual detections, thus suggesting
that we are not introducing significant bias by considering detections only.

\subsection{Excitation vs. Galaxy Properties}
\label{sec:results-galprops}

In the local universe, \citet{brinchmann2008} found that, at fixed stellar mass, 
galaxies with higher H$\alpha$ EWs (a proxy for sSFR) are offset towards
higher [NII]/H$\alpha$ and [OIII]/H$\beta$ in the [OIII]$\lambda 5007$/H$\beta$ vs. [NII]$\lambda 6585$/H$\alpha$ BPT diagram. \citet{liu2008} found that SDSS
galaxies with higher concentrations and $\Sigma_{SFR}$ values (again at fixed stellar mass)
tended to be offset in the same sense as high-redshift galaxies in the [OIII]$\lambda 5007$/H$\beta$ vs. [NII]$\lambda 6585$/H$\alpha$ BPT diagram.
To delve into the cause of anomalous properties of $z\sim 2.3$ star-forming
galaxies relative to local samples in the space of [OIII]$\lambda 5007$/H$\beta$ vs. [NII]$\lambda 6585$/H$\alpha$, we investigate the dependence of location in
the [OIII]$\lambda 5007$/H$\beta$ vs. [NII]$\lambda 6585$/H$\alpha$ diagram as a function of multiple different galaxy properties.
We separate the [OIII]$\lambda 5007$/H$\beta$ vs. [NII]$\lambda 6585$/H$\alpha$ MOSDEF sample into ``high" and ``low" bins
according to $M_*$, sSFR, and $\Sigma_{SFR}$, using the sample median of
each property ($M_{*,med}=10^{10.11} M_{\odot}$, sSFR$_{med}=2.4 \mbox{ Gyr}^{-1}$,
and $\Sigma_{SFR,med}=3.2 M_{\odot}\mbox{ yr}^{-1}\mbox{ kpc}^{-2}$) to demarcate the bins.
In Figure~\ref{fig:O3N2props}, we color-code $z\sim 2.3$ MOSDEF galaxies
in the [OIII]$\lambda 5007$/H$\beta$ vs. [NII]$\lambda 6585$/H$\alpha$ diagram by property, with high-mass [low-sSFR, low-$\Sigma_{SFR}$] points indicated in red,
and low-mass [high-sSFR, high-$\Sigma_{SFR}$] points in blue. Separate fits to the ``high"- and 
``low"-mass [sSFR, $\Sigma_{SFR}$] star-forming sequences are also indicated in each panel.

\begin{figure*}[t!]
\centerline{\includegraphics[height=3.5in]{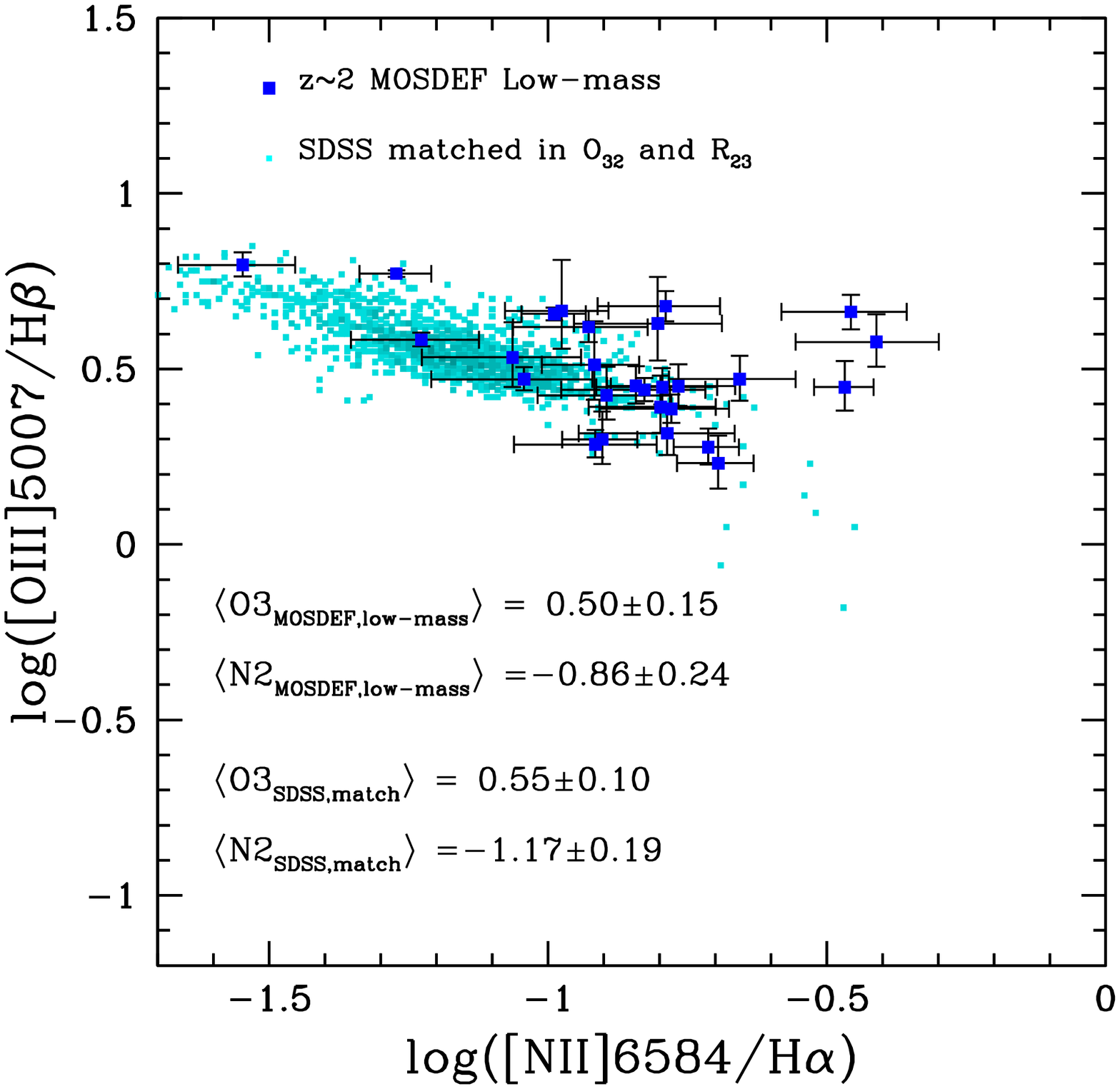}
\includegraphics[height=3.5in]{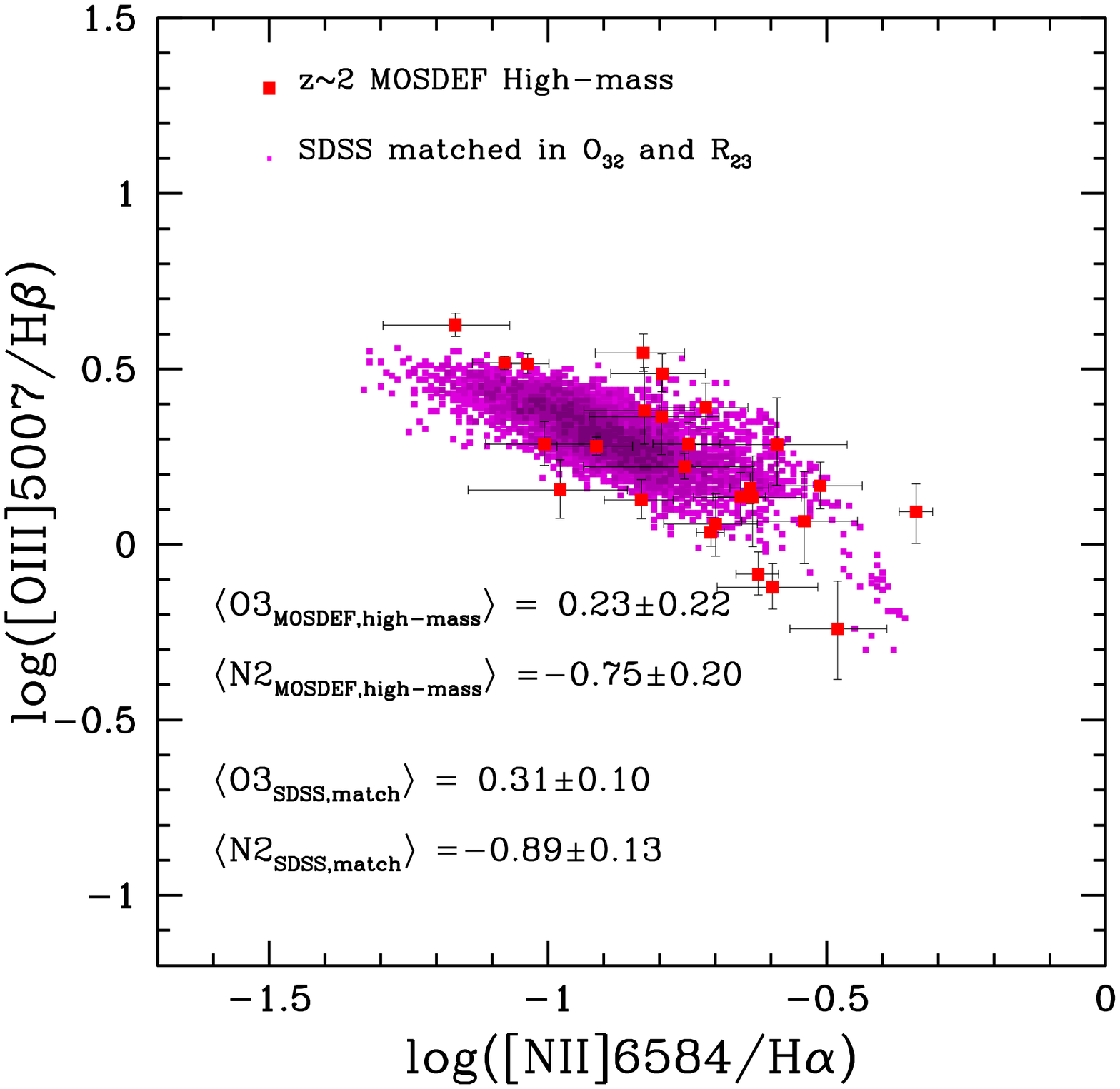}}
\caption{[OIII]$\lambda 5007$/H$\beta$ vs. [NII]$\lambda 6585$/H$\alpha$ BPT diagrams for low- and high-mass $z\sim 2.3$ MOSDEF galaxies, along with
matched SDSS samples. In each panel, the SDSS sample (small points) has been selected
to match the $z\sim 2.3$ MOSDEF subsample in average $\log(O_{32})$ and $\log(R_{23})$,
and the sample mean and standard deviation for O3 and N2 are listed for both MOSDEF and SDSS
samples. Given the MOSDEF and SDSS comparison sample sizes, the error on the mean
is a factor of $\sim 5$ smaller than the sample standard deviation for all MOSDEF samples,
and negligible for the SDSS comparison samples.
Left: Low-mass MOSDEF galaxies (blue points) and SDSS galaxies with matching
$O_{32}$ and $R_{23}$ (small cyan points) have very similar mean O3 ($\log(\mbox{[OIII]/H}\beta)$),
while N2 ($\log(\mbox{[NII]/H}\alpha)$) is offset towards higher values for the $z\sim 2.3$ MOSDEF galaxies.
Right: High-mass MOSDEF galaxies (red points) and SDSS galaxies with matching
$O_{32}$ and $R_{23}$ (small purple points) have very similar values for
both mean O3 and N2.}
\label{fig:o3n2-whatsgoingon}
\end{figure*}

The most striking segregation
perpendicular to the local excitation sequence is with stellar mass. The low-mass
MOSDEF galaxies occupy a region of [OIII]$\lambda 5007$/H$\beta$ vs. [NII]$\lambda 6585$/H$\alpha$ parameter space almost completely disjoint
from that of the SDSS sample. Furthermore, as shown in
Figure~\ref{fig:alldiagnostics-mass} (left), a fit to the star-forming sequence of the low-mass
MOSDEF galaxies is basically indistinguishable from the one
derived by \citet{steidel2014} for UV-selected galaxies.  At the same time, the 
high-mass half of the MOSDEF sample shows no significant offset relative to the SDSS 
sequence. Neither sSFR nor $\Sigma_{SFR}$ shows the same degree of separation
perpendicular to the SDSS star-forming sequence, in contrast to what is observed in the local
universe. With sSFR, galaxies separate
primarily along the dimension of [NII]/H$\alpha$, and, with
$\Sigma_{SFR}$, galaxies from the ``high" and ``low" samples are well-mixed in 
both dimensions. Due to the limited sample size of the early MOSDEF dataset, we are
not able to perform the same experiment that \citet{brinchmann2008} and \citet{liu2008}
did, looking for separation in sSFR and $\Sigma_{SFR}$ {\it at fixed stellar mass}. Such
analyses will be possible with the full MOSDEF dataset.

Given the clear separation according to stellar mass in the [OIII]$\lambda 5007$/H$\beta$ vs. [NII]$\lambda 6585$/H$\alpha$ diagram, we also investigate
how high- and low-mass MOSDEF galaxies populate the [OIII]$\lambda5007$/H$\beta$ vs. [SII]/H$\alpha$ and $O_{32}$ vs. $R_{23}$ diagrams 
(Figure~\ref{fig:alldiagnostics-mass}).
In all three panels, we color-code star-forming SDSS points by stellar mass. To
accentuate the fact that the mean stellar mass varies smoothly along the local star-forming sequences
in [OIII]$\lambda 5007$/H$\beta$ vs. [NII]$\lambda 6585$/H$\alpha$, [OIII]$\lambda5007$/H$\beta$ vs. [SII]/H$\alpha$, and $O_{32}$ vs. $R_{23}$, we only plot the low-mass ($M_* < 10^{9.25} M_{\odot}$; yellow) and high-mass
($M_*>10^{10.52} M_{\odot}$; green) tails of the SDSS stellar mass distribution.

In the [OIII]$\lambda5007$/H$\beta$ vs. [SII]/H$\alpha$ diagram, although low-mass galaxies have larger [OIII]/H$\beta$ ratios on average
than high-mass galaxies (by $\sim 0.3$~dex), both low- and high-mass samples show significant overlap
with the $z\sim 0$ star-forming sequence. Furthermore, given the shape of the $z\sim 0$
[OIII]$\lambda5007$/H$\beta$ vs. [SII]/H$\alpha$ star-forming sequence (i.e., more ``vertical" than the [OIII]$\lambda 5007$/H$\beta$ vs. [NII]$\lambda 6585$/H$\alpha$ sequence), the separation by mass 
at $z\sim 2.3$ is not obviously inconsistent with the manner in which local galaxies separate by  mass
(i.e., low-mass galaxies primarily show higher [OIII]/H$\beta$, although the low- and high-mass samples significantly
overlap in [SII]/H$\alpha$ as the star-forming sequence curves back on itself). In contrast,
the separation of low- and high-mass MOSDEF galaxies in the [OIII]$\lambda 5007$/H$\beta$ vs. [NII]$\lambda 6585$/H$\alpha$ diagram extends in a
direction almost perpendicular to the sense in which low- and high-mass galaxies segregate along the 
local [OIII]$\lambda 5007$/H$\beta$ vs. [NII]$\lambda 6585$/H$\alpha$ sequence. 

In the $O_{32}$ vs. $R_{23}$ panel, the separation of low- and high-mass MOSDEF galaxies is entirely
parallel to the local star-forming sequence and the sense in which local galaxies segregate, 
with low-mass objects showing both significantly higher $O_{32}$ values and slightly higher values of $R_{23}$. 
One final point is that, in all panels, the MOSDEF sample is entirely contained within the region dominated
by low-mass ($M_* < 10^{9.25} M_{\odot}$) SDSS galaxies -- i.e., the $z\sim 2.3$ sample shows excitation 
properties characteristic of low-mass galaxies in the local universe. We return to this point
in section~\ref{sec:discussion}.

\begin{figure*}
\centerline{\includegraphics[height=3.5in]{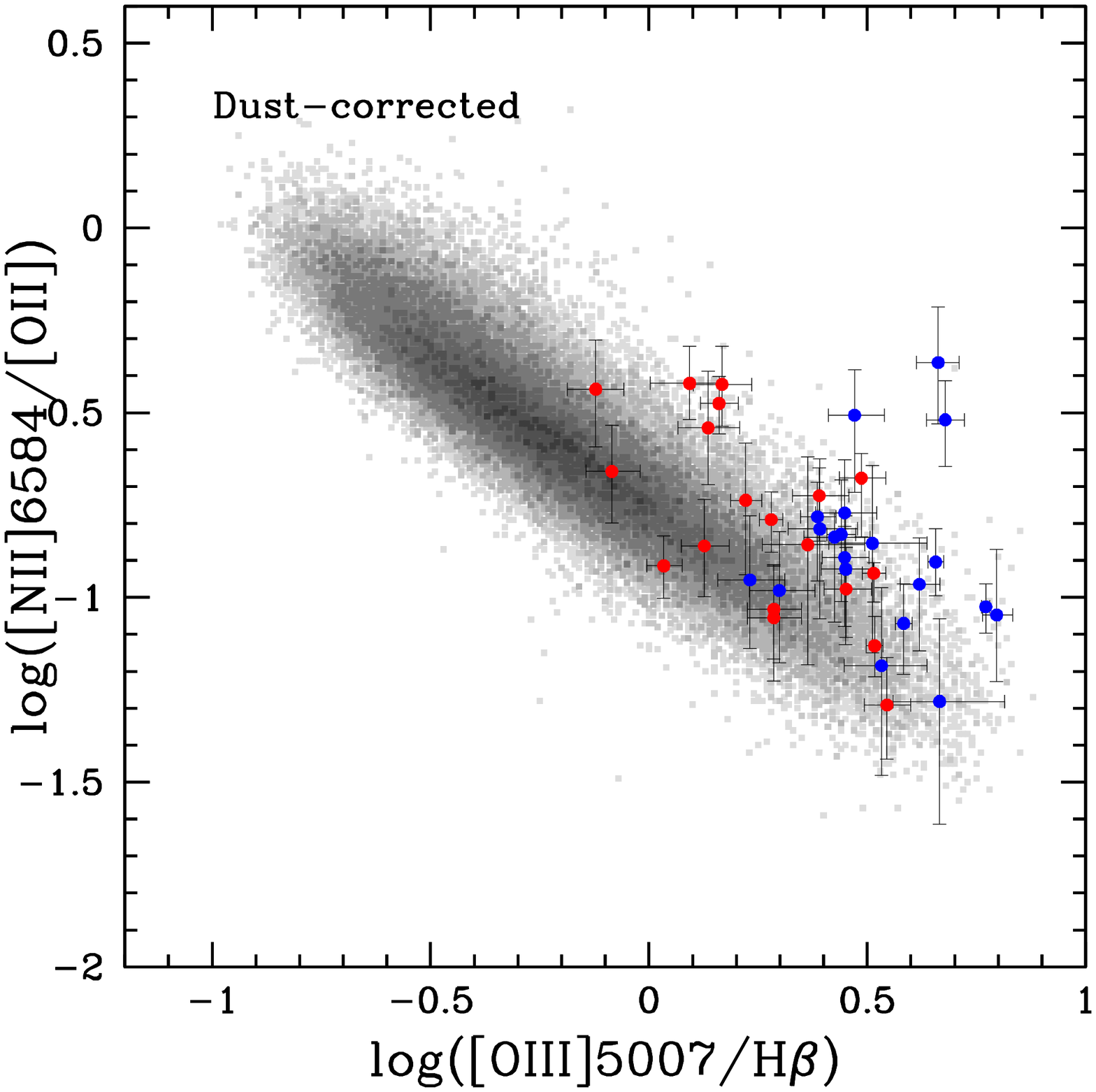}\includegraphics[height=3.5in]{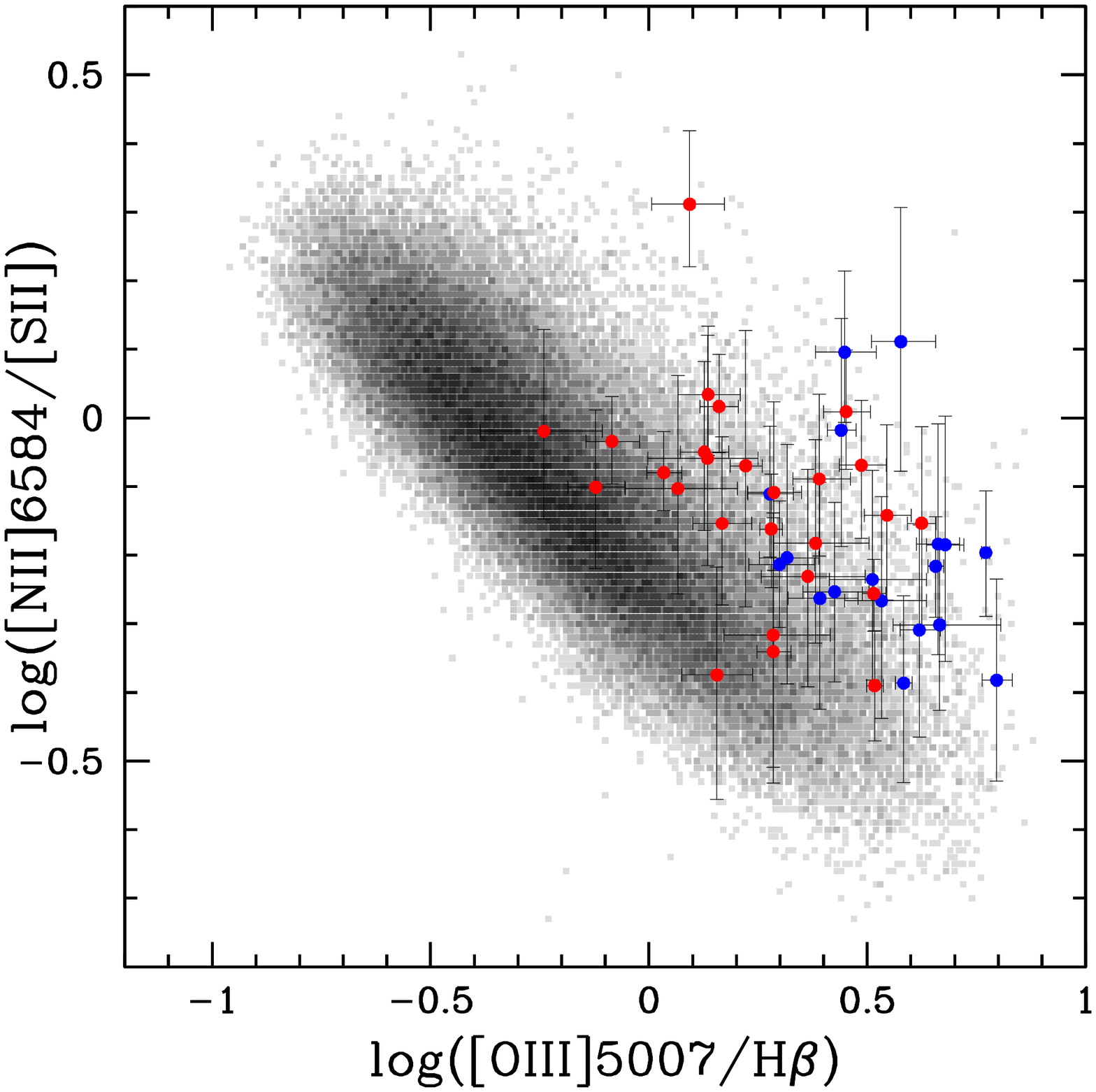}}
\caption{Left: N2O2 vs. [OIII]/H$\beta$. Symbols are as in Figure~\ref{fig:O3N2props}, with
low- and high-mass $z\sim 2.3$ MOSDEF galaxies indicated, respectively, with blue and
red points. N2O2 values
have been corrected for dust extinction, given the wide wavelength baseline between [NII]$\lambda6584$
and [OII]$\lambda\lambda3726,3729$. Among the low-mass $z\sim 2.3$ MOSDEF galaxies (those
at the highest [OIII]/H$\beta$ values), there is an offset towards higher N2O2 at fixed
[OIII]/H$\beta$, which we interpret as an elevated N/O abundance ratio at fixed oxygen
abundance. At higher masses in the $z\sim 2.3$ sample, there is less evidence for an elevated N/O ratio.
Right: N2S2 vs. [OIII]/H$\beta$. Again, symbols are as in Figure~\ref{fig:O3N2props}.
Independent of stellar mass, $z\sim 2.3$ MOSDEF galaxies are offset towards higher N2S2 values
at fixed [OIII]/H$\beta$. This trend may indicate evidence for variations in S/O ratios
when considered together with the N2O2 vs. [OIII]/H$\beta$ diagram.}
\label{fig:noo3nso3}
\end{figure*}

\subsection{An offset in N2 or O3?}
\label{sec:results-o3orn2}

In order to understand the underlying cause of the offset in the [OIII]$\lambda 5007$/H$\beta$ vs. [NII]$\lambda 6585$/H$\alpha$ BPT diagram,
it is important to determine the manner in which the high-redshift systems
are actually different from local galaxies. Specifically, we would like to distinguish
among the possibilities that galaxies with a given set of physical properties (e.g., metallicity)
have higher [OIII]/H$\beta$ ratios (O3), higher [NII]/H$\alpha$ ratios (N2), or both higher 
O3 and N2 values.  MOSDEF measurements in multiple diagnostic diagrams enable this
investigation. Figures~\ref{fig:O3N2O3S2}
and \ref{fig:O32R23} already suggest that emission diagnostics involving nitrogen are the source
of the anomaly. With the use of the $O_{32}$ vs. $R_{23}$ diagram, we can provide a definitive demonstration that this
is the case. In the $O_{32}$ vs. $R_{23}$ diagram, $z\sim 2.3$ MOSDEF galaxies closely follow the distribution of the local SDSS
sample, albeit occupying the high-$O_{32}$, high-$R_{23}$ portion of the local excitation sequence. 
Accordingly, it is possible to construct a sample of SDSS galaxies that matches a given set of 
MOSDEF galaxies in mean $O_{32}$ and $R_{23}$, and then compare the distributions of O3 and N2 for 
both the MOSDEF and matched SDSS samples.
We construct matched SDSS samples for both low-mass ($M_* < 10^{10.11} M_{\odot}$) and high-mass
($M_* \geq 10^{10.11} M_{\odot}$) MOSDEF galaxies and plot the matched MOSDEF and SDSS samples
in the space of [OIII]$\lambda 5007$/H$\beta$ vs. [NII]$\lambda 6585$/H$\alpha$ (Figure~\ref{fig:o3n2-whatsgoingon}), along with an indication of the sample
mean O3 and N2 values. The high-mass MOSDEF and matched SDSS
samples follow very similar distributions in [OIII]$\lambda 5007$/H$\beta$ vs. [NII]$\lambda 6585$/H$\alpha$, with similar mean O3 and N2 values within
the errors. Although
the low-mass MOSDEF and matched SDSS sample have very similar mean O3 values, the low-mass MOSDEF
sample is offset towards higher N2 values. This analysis demonstrates that the offset in the $z\sim 2.3$
[OIII]$\lambda 5007$/H$\beta$ vs. [NII]$\lambda 6585$/H$\alpha$ BPT diagram should be understood in terms of low-mass $z\sim 2$ MOSDEF galaxies having higher 
[NII]/H$\alpha$ ratios on average, at fixed [OIII]/H$\beta$. Alternatively, we can state that, 
at fixed $O_{32}$ and $R_{23}$, which, as we argue in section~\ref{sec:discussion-o32r23}, serves as a
proxy for oxygen abundance, low-mass $z\sim 2.3$ MOSDEF galaxies have higher [NII]/H$\alpha$ ratios
than their $z\sim 0$ counterparts. 

\section{Discussion}
\label{sec:discussion}

\subsection{The [NII]/H$\alpha$ Offset in the [OIII]$\lambda 5007$/H$\beta$ vs. [NII]$\lambda 6585$/H$\alpha$ BPT Diagram}
\label{sec:discussion-n2offset}

There are several different effects that have been considered to explain the observed offset
in the [OIII]$\lambda 5007$/H$\beta$ vs. [NII]$\lambda 6585$/H$\alpha$ BPT diagram among high-redshift galaxies. First, there are the physical parameters
describing the H~II regions contributing to the integrated line ratios from galaxies. These
include the ionization parameter, ionizing spectrum of the stars illuminating the H~II-region
gas, and the electron density. Early work highlighting the issue of the high-redshift offset
in the [OIII]$\lambda 5007$/H$\beta$ vs. [NII]$\lambda 6585$/H$\alpha$ BPT diagram
focused on these parameters \citep[e.g.,][]{shapley2005b,liu2008,brinchmann2008},
and they have been revisited more recently by \citet{kewley2013}, \citet{steidel2014}, and \citet{masters2014}. 
Systematically higher ionization parameters \citep[which appear to apply in high-redshift
galaxies;][]{nakajima2013}, harder ionizing spectra, and higher electron densities
\citep{shirazi2014}, all tend to shift the locus of galaxies in the [OIII]$\lambda 5007$/H$\beta$ vs. [NII]$\lambda 6585$/H$\alpha$ diagram
towards higher [OIII]/H$\beta$ and [NII]/H$\alpha$ values.
Next, there is the role of different types of pressure in determining the internal
structure and dynamics of H~II regions. \citet{yeh2013} and \citet{verdolini2013}
suggest that radiation pressure is significant in high-redshift H~II regions, as
compared with gas pressure associated with stellar winds, and that the effects
of radiation pressure can lead to [OIII]/H$\beta$ line ratios in excess of the 
``maximum starburst" limit of \citet{kewley2001}. Possible contamination by
weak AGNs has also been suggested as a way to shift galaxy emission-line
ratios into the ``composite" region of the [OIII]$\lambda 5007$/H$\beta$ vs. [NII]$\lambda 6585$/H$\alpha$ BPT diagram \citep[e.g.,][]{wright2010}, in between
the curves of \citet{kauffmann2003} and \citet{kewley2001}. Finally, both
\citet{masters2014} and \citet{steidel2014} consider
gas-phase abundance ratios -- specifically, the N/O ratio -- which
can affect where galaxies fall in the [OIII]$\lambda 5007$/H$\beta$ vs. [NII]$\lambda 6585$/H$\alpha$ BPT diagram. 
If the relationship between N/O and $12+\log(\mbox{O/H})$ evolves out to 
high redshift, then distant galaxies will shift relative to local ones 
in the [OIII]$\lambda 5007$/H$\beta$ vs. [NII]$\lambda 6585$/H$\alpha$ BPT diagram.

In order to distinguish among these different scenarios, it is crucial to consider
the full set of strong rest-frame optical emission lines. These include not only
H$\beta$, [OIII]$\lambda5007$, H$\alpha$, and [NII]$\lambda6584$, which comprise
the [OIII]$\lambda 5007$/H$\beta$ vs. [NII]$\lambda 6585$/H$\alpha$ BPT measurements, but also [OII]$\lambda\lambda3726,3729$ and [SII]$\lambda,\lambda6717,6731$.
With the $z\sim 2.3$ MOSDEF dataset, we are in the unique position of covering
this full set of lines for a statistical sample of galaxies. Accordingly, we investigate both
[OIII]$\lambda 5007$/H$\beta$ vs. [NII]$\lambda 6585$/H$\alpha$ and [OIII]$\lambda5007$/H$\beta$ vs. [SII]/H$\alpha$ BPT diagrams, as well as the space of $O_{32}$ vs. $R_{23}$. Based on a single composite spectrum
of 26 emission-line selected galaxies at $1.3 \leq z \leq 2.3$, \citet{masters2014}
found that the observed offset in the [OIII]$\lambda 5007$/H$\beta$ vs. [NII]$\lambda 6585$/H$\alpha$ BPT diagram did not appear in the space of [OIII]$\lambda5007$/H$\beta$ vs. [SII]/H$\alpha$.
We confirm this same trend, using a larger sample of individual measurements. Specifically,
the MOSDEF sample scatters symmetrically around the $z\sim 0$ star-forming sequence
in the [OIII]$\lambda5007$/H$\beta$ vs. [SII]/H$\alpha$ BPT diagram, and does not suggest a systematic offset. Even more
striking is the distribution of MOSDEF galaxies in the $O_{32}$ vs. $R_{23}$ diagram, which is completely based
on oxygen and hydrogen emission lines. In this space, the $z\sim 2.3$ MOSDEF sample
scatters along the local sequence, with no evidence for an offset perpendicular
to the locus of $z\sim 0$ galaxies. Although the $z\sim 2.3$ galaxies only occupy
the high-$O_{32}$, high-$R_{23}$ tail of the local distribution, we can identify
direct analogs for high-redshift galaxies in the space of $O_{32}$ vs. $R_{23}$. We use this
match to uncover the nature of the [OIII]$\lambda 5007$/H$\beta$ vs. [NII]$\lambda 6585$/H$\alpha$ offset among low-mass $z\sim 2.3$ MOSDEF
galaxies, finding that SDSS galaxies matched to $z\sim 2.3$ samples in average $O_{32}$ and $R_{23}$
are also roughly matched in average [OIII]/H$\beta$. The offset for these low-mass $z\sim 2.3$
MOSDEF galaxies is found in their [NII]/H$\alpha$ ratios. Indeed, at fixed $O_{32}$, $R_{23}$,
and [OIII]/H$\beta$, the $z\sim 2.3$ galaxies offset in the [OIII]$\lambda 5007$/H$\beta$ vs. [NII]$\lambda 6585$/H$\alpha$ BPT diagram show higher
average [NII]/H$\alpha$ ratios than their SDSS counterparts.

As in \citet{masters2014}, we interpret this offset as a difference in N/O abundance
ratio at fixed metallicity. We find further evidence of a deviation in N/O when
considering the [NII]$\lambda6584$/[OII]$\lambda\lambda3726,3729$ ratio (N2O2).
N2O2  has been calibrated by \citet{perezmontero2009}
for local galaxies as a proxy for the N/O abundance ratio ($\log(\mbox{N/O})=0.93\times\mbox{N2O2}-0.20$). 
In order to compare with
SDSS, we plot N2O2 as a function of [OIII]$\lambda5007$/H$\beta$. Since the set
of excitation diagrams we have considered here collectively suggest that [NII]/H$\alpha$
is the anomalous line ratio at high redshift, [OIII]$\lambda5007$/H$\beta$
serves as a better control variable than [NII]/H$\alpha$, and is
anti-correlated with oxygen abundance in the regime that we consider
here \citep{maiolino2008}. For the low-mass portion of the MOSDEF sample,
corresponding to the highest [OIII]/H$\beta$ ratios ($\langle\log(\mbox{[OIII]/H}\beta)\rangle=0.5$),
Figure~\ref{fig:noo3nso3} (left) demonstrates that MOSDEF galaxies are characterized
by higher N2O2 ratios on average relative to SDSS galaxies with similar [OIII]/H$\beta$ ratios.
This average offset corresponds to $\Delta$N2O2=0.24 dex.
For the high-mass portion of the MOSDEF sample (at lower [OIII]/H$\beta$ ratios), 
there is better agreement between $z\sim 2.3$ MOSDEF and SDSS galaxies. This result
provides additional support for higher N/O ratios at fixed oxygen abundance
among $z\sim 2.3$ MOSDEF galaxies with $M\leq 10^{10.11} M_{\odot}$. 

\citet{perezmontero2009}
also provide a calibration between [NII]$\lambda6584$/[SII]$\lambda\lambda6717,6731$ ratio (N2S2)
and N/O ($\log(\mbox{N/O})=1.26\times\mbox{N2S2}-0.86$). This relationship is characterized
by more scatter than the one using N2O2, possibly due
to variations in the S/O abundance ratio in galaxies. We show the relation between
N2S2 and [OIII]/H$\beta$ for $z\sim 2.3$ MOSDEF and SDSS galaxies in Figure~\ref{fig:noo3nso3} (right),
uncovering an offset towards higher N2S2 at fixed [OIII]/H$\beta$, independent of [OIII]/H$\beta$
(i.e., mass). Exactly why the behavior of N2S2 and N2O2 differ in detail at high mass is
beyond the scope of the current work, as is detailed photoionization modeling
to constrain the nature of high-redshift ionizing spectra and ionization parameters,
and the significance of radiation
pressure in determining H~II region structure and dynamics. 
However, with our unique, multi-dimensional dataset, we establish that, at low-mass, $z\sim 2.3$
MOSDEF galaxies consistently show an offset towards higher N/O ratios at fixed [OIII]/H$\beta$,
relative to the SDSS sample. In order to investigate potential differences 
in ionizing spectra and ionization parameters at high redshift, photoionization models
tuned to match the locus of high-redshift galaxies in the [OIII]$\lambda 5007$/H$\beta$ vs. [NII]$\lambda 6585$/H$\alpha$ diagram alone \citep[e.g.,][]{steidel2014} 
must now be compared with observations of the [OIII]$\lambda5007$/H$\beta$ vs. [SII]/H$\alpha$ and $O_{32}$ vs. $R_{23}$ diagrams. Such comparisons
are required to test the general validity of these models.

\begin{figure*}
\centerline{\includegraphics[height=3.5in]{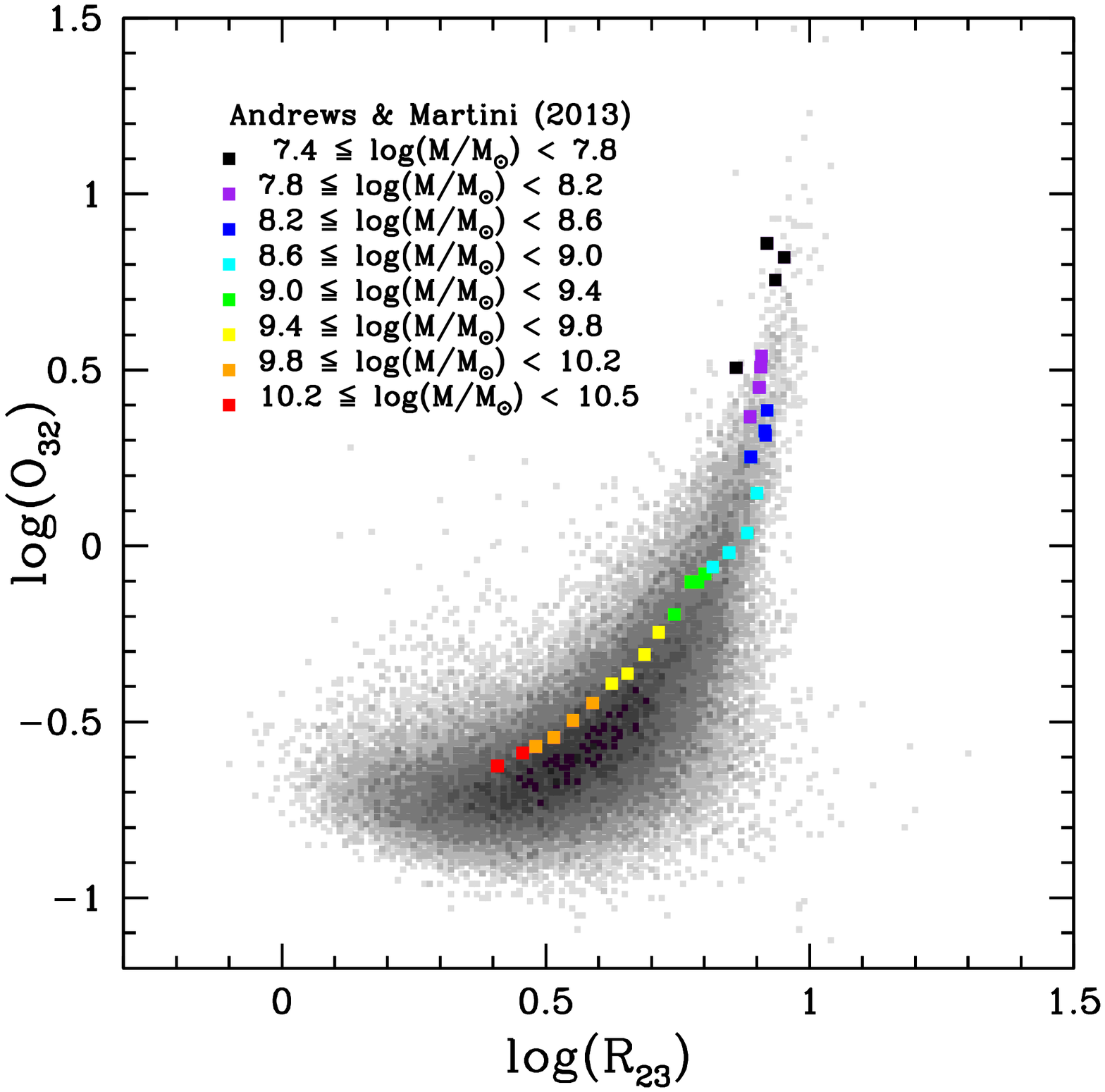}\includegraphics[height=3.5in]{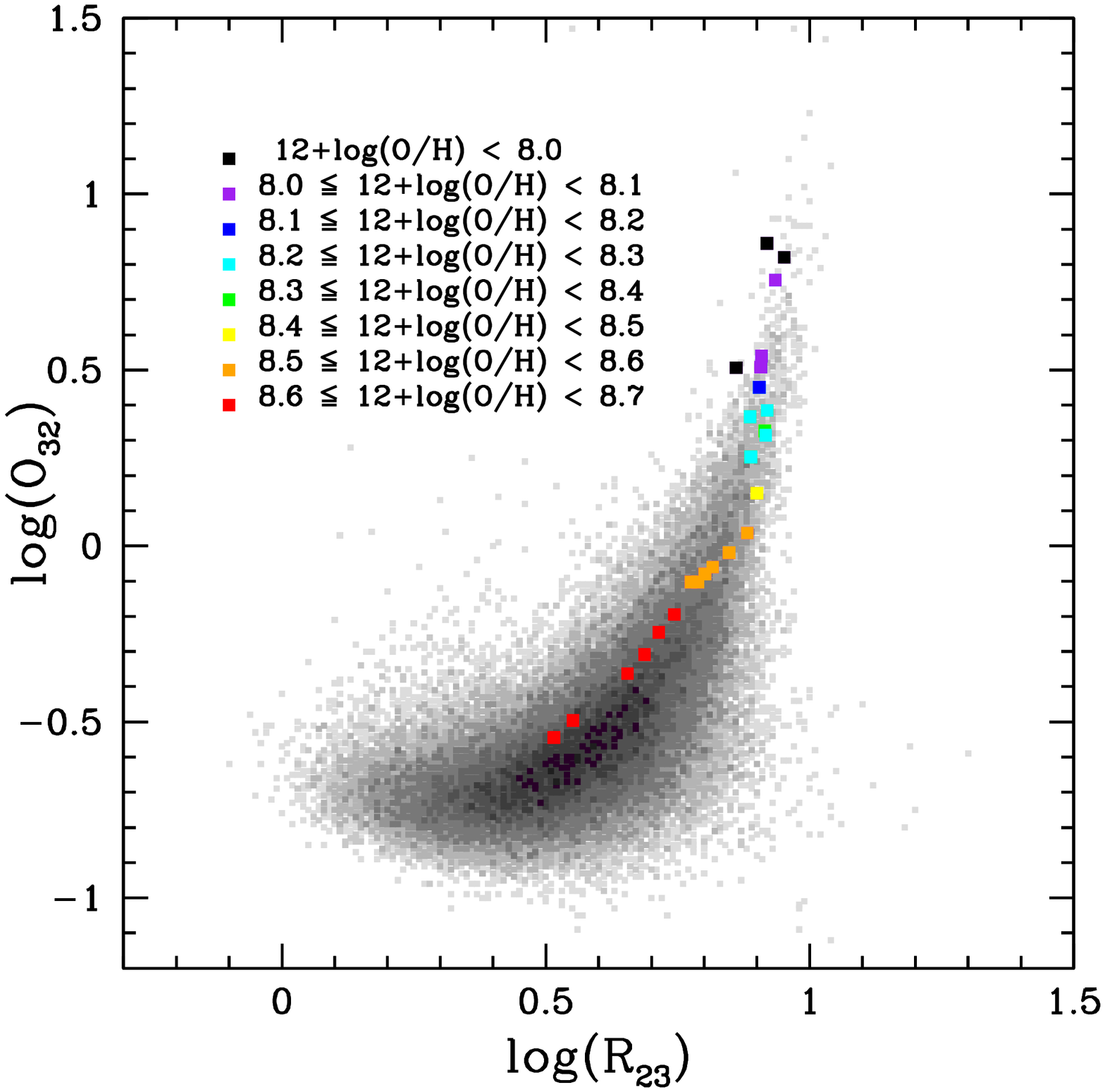}}
\caption{SDSS $O_{32}$ vs. $R_{23}$ excitation diagram, along with measurements from 
SDSS composite spectra from \citet{andrews2013} with direct oxygen abundances.
In both panels, points have been corrected for dust extinction, and the larger 
symbols represent strong-line ratios measured from the \citeauthor{andrews2013} composite spectra in bins 
of stellar mass. Left: The legend indicates how color corresponds to stellar mass for each composite
spectrum.  As in both the samples of individual SDSS and $z\sim 2.3$
MOSDEF galaxies, the \citeauthor{andrews2013} composites indicate a clear progression
in stellar mass along the $O_{32}$ vs. $R_{23}$ excitation sequence. Right: The
legend indicates how color corresponds to metallicity for each composite spectrum.
The metallicity of the \citeauthor{andrews2013}
stacks decreases monotonically along the $O_{32}$ vs. $R_{23}$ sequence, as $O_{32}$ and $R_{23}$ increase.
This smooth decline in metallicity suggests the combination of $O_{32}$ and $R_{23}$
as a powerful metallicity indicator, which can be calibrated to the 
direct oxygen abundance scale.}
\label{fig:o32r23-andmar13}
\end{figure*}

\citet{masters2014} and \citet{steidel2014} offer differing explanations for elevated
N/O ratios in high-redshift, actively star-forming galaxies. \citet{masters2014}
attribute this abundance signature to enhanced populations of Wolf-Rayet stars
in $z\sim 2$ stellar populations, whose nitrogen-rich winds (from WN-sequence Wolf-Rayet
stars) mix with the ISM to enrich future generations of stars. Such a scenario 
relies on catching galaxies at extremely young stellar population ages, when
the signatures of Wolf-Rayet stars are most prominent in integrated spectra \citep{shapley2003}. 
This timing requirement may prove unrealistic, depending on how common the elevated
N/O ratios prove to be with a larger sample. \citet{steidel2014}
on the other hand appeal to stellar population models including binaries and rotation
in massive stars \citep[e.g.,][]{eldridge2009}. Such stellar evolution models
may be particularly appropriate for high-redshift
star-forming galaxies, as argued by \citet{steidel2014}, and feature enhanced production
of nitrogen by massive stars during their main sequence evolution. Stellar population
synthesis models and observations
of the rest-frame UV spectra of galaxies over a wide range of N/O abundance
ratios will prove a promising avenue for connecting observations of 
rest-frame optical emission lines and high-redshift stellar populations. The rest-frame
UV contains multiple features associated with stellar winds (e.g., C~IV$\lambda\lambda1548,1550$
P-Cygni and He~II$\lambda1640$), and a correlation of the properties of these features
with N/O will likely provide important constraints on chemical enrichment 
and stellar evolution in the early universe.

\subsection{The $O_{32}$ vs. $R_{23}$ Diagram: A Robust Abundance Indicator at High Redshift}
\label{sec:discussion-o32r23}

The offset in the [OIII]$\lambda 5007$/H$\beta$ vs. [NII]$\lambda 6585$/H$\alpha$ BPT diagram raises concerns about estimating
metallicities at high redshift from strong-line methods that include
nitrogen features \citep[e.g., the $N2$ and $O3N2$ indicators;][]{pp2004}.
Various authors have attempted to quantify the bias in applying the locally-calibrated
$N2$ and $O3N2$ indicators to high-redshift samples that follow different distributions
in the [OIII]$\lambda 5007$/H$\beta$ vs. [NII]$\lambda 6585$/H$\alpha$ BPT diagram \citep{liu2008,newman2014,steidel2014}. Both \citet{liu2008}
and \citet{newman2014} conclude that the N2 indicator will overestimate gas-phase
oxygen abundances for high-redshift galaxies, while \citet{liu2008}
and \cite{steidel2014} conclude that the $O3N2$ indicator should not be significantly
biased, relative to its use in the local universe. However, this indicator still
includes the [NII]/H$\alpha$ ratio, the value of which relative to oxygen abundance will depend on the 
N/O ratio. 

In the $O_{32}$ vs. $R_{23}$ diagram, on the other hand, there is no observed
offset between the $z\sim 2.3$ MOSDEF sample and local galaxies. Furthermore, the emission
lines featured in this diagram consist only of oxygen and hydrogen transitions, i.e.
more direct probes of the oxygen abundance, and are immune to the variations in abundance
ratios that affect abundance indicators based on nitrogen lines. These factors suggest
that an abundance indicator based on $O_{32}$ and $R_{23}$ holds much promise
for tracking the evolution in oxygen abundance over cosmic time. Although the $R_{23}$ indicator
has commonly been used to estimate oxygen abundances, \citet{steidel2014} points out that
it is of limited sensitivity over the range of metallicities spanned by the majority of 
high-redshift galaxies studied thus far (see their Figure 10). We confirm this result,
in that $\sim 90$\% of the $z\sim 2.3$ MOSDEF galaxies with $R_{23}$ measurements fall within
a range of only 0.4~dex in $R_{23}$. However, we observe a significant spread of the sample
when considering the combination of $R_{23}$ and $O_{32}$, in that
$O_{32}$ increases over a range of $\sim 1.5$~dex as $R_{23}$ increases by $\sim 0.5$~dex.
This same trend is traced out with much more significance in the SDSS sample, where the full
sequence in $O_{32}$ and $R_{23}$ extends from a plateau of roughly constant $\log(O_{32})$
values of $\sim -0.7$~dex and $\log(R_{23})=0.1-0.5$~dex, up through a correlated region
where $\log(O_{32})$ and $\log(R_{23})$ both increase up to $\sim 1$~dex.

The translation of this sequence to one of physical properties becomes clear
as we consider the segregation along it with stellar mass. Figure~\ref{fig:alldiagnostics-mass}
shows a clean separation in the $z\sim 2.3$ MOSDEF sample at the median stellar mass, with 
low-mass galaxies occupying the upper-right portion of the $z\sim 2.3$ $O_{32}$ vs. $R_{23}$ distribution, 
and high-mass galaxies the lower-left. The SDSS sample also shows a systematic variation
of mass with position along the sequence of $O_{32}$ vs. $R_{23}$, 
with the lowest-mass systems dominating the distribution at the highest $O_{32}$ and $R_{23}$ values 
and the highest-mass galaxies dominating the plateau at low $O_{32}$ and $R_{23}$.
Given the strong correlation between stellar mass and oxygen abundance, this segregation by 
mass is a direct reflection of metallicity decreasing systematically along the $O_{32}$ vs $R_{23}$ 
sequence, as we traverse from the high-mass, low $O_{32}$ and $R_{23}$ regime into the one of
the low-mass, high $O_{32}$ and $R_{23}$. 

Figure~\ref{fig:alldiagnostics-mass} (right) in fact
reflects the well-known evolution in the mass-metallicity relation (MZR) observed between 
$z\sim 0$ and $z\sim 2$ \citep[e.g.,][]{erb2006a,sanders2014,steidel2014}. Accordingly,
at fixed stellar mass, galaxies appear to evolve by somewhere between $\sim 0.15$ and $0.3$~dex
towards lower metallicities (though the exact evolution inferred depends on how metallicities
are estimated and the high-redshift sample used for comparison with SDSS). This evolution 
can also be cast in terms of galaxies of fixed metallicity
corresponding to lower-mass galaxies in the local universe than at high redshift. 
The median stellar mass of the $z\sim 2.3$ MOSDEF sample plotted in Figure~\ref{fig:alldiagnostics-mass}
is $\sim 10^{10} M_{\odot}$, while the median stellar mass of the SDSS galaxies occupying
the same portion of the $O_{32}$ vs. $R_{23}$ diagram is $\sim 10^9 M_{\odot}$. Alternatively,
when we construct an SDSS sample matched in median stellar mass to the $z\sim 2.3$ MOSDEF
sample, we find that it is offset towards lower $O_{32}$ and $R_{23}$ values relative
to the $z\sim 2.3$ MOSDEF sample. These two results follow directly if location along the
sequence of $O_{32}$ vs. $R_{23}$ corresponds closely to oxygen abundance.

We use the composite spectra of \citet{andrews2013} to demonstrate the utility
of the $O_{32}$ vs. $R_{23}$ diagram as a metallicity indicator. \citet{andrews2013} constructed
composite spectra for SDSS galaxies in bins of stellar mass, and used 
measurements of auroral [OII] and [OIII]
emission lines to estimate both electron temperatures and 
``direct" oxygen abundances as a function of stellar mass. Such estimates of $12+\log(\mbox{O/H})$
are completely independent of the strong-line methods commonly used to determine
metallicity, and therefore offer a powerful empirical probe of the connection between metallicity
and the observed patterns of strong emission lines. In Figure~\ref{fig:o32r23-andmar13} (left),
we show a clear progression in stellar mass among the \citeauthor{andrews2013} composite spectra
along the $O_{32}$ vs. $R_{23}$ sequence. Again, lower-mass composites in the \citeauthor{andrews2013}
sample occupy the upper right portion of the $O_{32}$ vs. $R_{23}$ sequence, while
the higher-mass composites curve down and to the left.  
Since the majority of stellar mass composites from \citet{andrews2013}
have direct oxygen abundances, we can also examine how direct oxygen abundance varies as a
function of location along the sequence of $O_{32}$ vs. $R_{23}$. Figure~\ref{fig:o32r23-andmar13}
(right) shows that there is a basically monotonic decrease in metallicity along 
the $O_{32}$ vs. $R_{23}$ sequence, extending from the lower left to upper right.
This clean variation of direct oxygen abundance with location in the $O_{32}$ vs. $R_{23}$ diagram
suggests an extremely promising metallicity indicator. Given that the $z\sim 2.3$ MOSDEF sample
follows the same distribution as SDSS galaxies in $O_{32}$ vs. $R_{23}$, a calibration
of oxygen abundance with $O_{32}$ vs. $R_{23}$ location based on SDSS should also work for $z\sim 2$.

The $O_{32}$ vs. $R_{23}$ diagram has long been known as a probe of both ionization parameter and
metallicity \citep{kewley2002}, and has even been used to estimate oxygen
abundances at $z\sim 2$ \citep{nakajima2013,nakajima2014}. In fact, just as we discuss above,
\citet{nakajima2013} also point out the fact that low-mass, low-metallicity
SDSS galaxies occupy the same region of $O_{32}$ vs. $R_{23}$ parameter space
as high-redshift galaxies. However, in both \citet{kewley2002}
and \citet{nakajima2013}, the combination of $O_{32}$ and $R_{23}$ is translated
to a combination of metallicity and ionization parameter using the results of photoionization
models. In contrast, here we propose to estimate oxygen abundances with an empirical calibration of
$O_{32}$ and $R_{23}$ values based on direct oxygen abundances (Shapley et al., in prep). 
This  $O32R23$ abundance indicator will be
tied to the abundance scale of the direct oxygen abundances and applicable out
to at least $z\sim 2$, given the apparent lack of evolution in the $O_{32}$ vs. $R_{23}$ sequence over
this interval. Accordingly, it will be possible to infer metallicities in an unbiased
fashion over a wide range of redshifts with the measurement of dust-corrected
[OII]$\lambda\lambda 3726,3729$, H$\beta$, and [OIII]$\lambda5007$ emission fluxes.
Dust corrections ideally would be obtained using measurements of the Balmer decrement
from the H$\alpha$/H$\beta$ ratio (Reddy et al., in prep.), but may also be estimated from stellar population modeling,
with the assumption of how nebular and stellar extinctions are related \citep[e.g.,][]{erb2006c,
forsterschreiber2009,steidel2014}. 

The simultaneous detection of [OII]$\lambda\lambda 3726,3729$, H$\beta$, [OIII]$\lambda5007$,
and H$\alpha$ emission lines is becoming feasible at $z>1$ with the current generation of
multi-object near-IR spectrographs. However, given that most high-redshift datasets
are still based on more limited sets of rest-frame optical emission lines, 
it will be important to determine
the translations between oxygen abundances based on commonly-used indicators such as $N2$ and 
$O3N2$, and the directly-calibrated $O32R23$ indicator proposed here. Such translations would
be analogous to those proposed by \citet{kewley2008} for the local universe, but
appropriate for higher-redshift samples with different footprints in the 
[OIII]/H$\beta$ vs. [NII]/H$\alpha$ BPT diagram \citep[see][for a discussion of
how the translation between the $N2$ and $O3N2$ indicators
evolves with redshift]{sanders2014,steidel2014}.

With unbiased oxygen abundance measurements in hand, it will be possible
to determine the evolution of the shape and normalization of the MZR in a robust fashion
over at least 10 Gyr, with important implications for understanding the nature of 
gas flows in galaxies \citep{finlator2008,dave2011, dave2012}.
We must also understand the origins of the differences in N/O
abundance ratios at high redshift as well as characterize the physical conditions
in high-redshift star-forming regions. The rich dataset of the MOSDEF survey will
enable such investigations in an unprecedented manner.

\acknowledgements
We thank the referee for an extremely constructive report.
We acknowledge support from NSF AAG grants AST-1312780, 1312547, 1312764, and 1313171 and grant AR-13907
from the Space Telescope Science Institute. We are
also grateful to Marc Kassis at the Keck Observatory for his many valuable contributions to
the execution of the MOSDEF survey. We also acknowledge
the 3D-HST collaboration, who provided us with spectroscopic
and photometric catalogs used to select MOSDEF targets and derive
stellar population parameters. We thank I. McLean, K. Kulas,
and G. Mace for taking observations for the MOSDEF survey in May and June 2013.
MK acknowledges support from a Committee Faculty Research Grant and a Hellman Fellowship. ALC
acknowledges funding from NSF CAREER grant AST-1055081. NAR is supported by
an Alfred P. Sloan Research Fellowship.
We wish to extend special thanks to those of Hawaiian ancestry on
whose sacred mountain we are privileged to be guests. Without their generous hospitality, most
of the observations presented herein would not have been possible.

\bibliographystyle{apj}
\bibliography{mosdef}

\begin{thebibliography}{}
\expandafter\ifx\csname natexlab\endcsname\relax\def\natexlab#1{#1}\fi

\bibitem[{{Abazajian} {et~al.}(2009){Abazajian}, {Adelman-McCarthy},
  {Ag{\"u}eros}, {Allam}, {Allende Prieto}, {An}, {Anderson}, {Anderson},
  {Annis}, {Bahcall}, \& et~al.}]{abazajian2009}
{Abazajian}, K.~N., {Adelman-McCarthy}, J.~K., {Ag{\"u}eros}, M.~A., {et~al.}
  2009, \apjs, 182, 543

\bibitem[{{Andrews} \& {Martini}(2013)}]{andrews2013}
{Andrews}, B.~H., \& {Martini}, P. 2013, \apj, 765, 140

\bibitem[{{Baldwin} {et~al.}(1981){Baldwin}, {Phillips}, \&
  {Terlevich}}]{baldwin1981}
{Baldwin}, J.~A., {Phillips}, M.~M., \& {Terlevich}, R. 1981, \pasp, 93, 5

\bibitem[{{Brammer} {et~al.}(2012){Brammer}, {van Dokkum}, {Franx},
  {Fumagalli}, {Patel}, {Rix}, {Skelton}, {Kriek}, {Nelson}, {Schmidt},
  {Bezanson}, {da Cunha}, {Erb}, {Fan}, {F{\"o}rster Schreiber}, {Illingworth},
  {Labb{\'e}}, {Leja}, {Lundgren}, {Magee}, {Marchesini}, {McCarthy},
  {Momcheva}, {Muzzin}, {Quadri}, {Steidel}, {Tal}, {Wake}, {Whitaker}, \&
  {Williams}}]{brammer2012}
{Brammer}, G.~B., {van Dokkum}, P.~G., {Franx}, M., {et~al.} 2012, \apjs, 200,
  13

\bibitem[{{Brinchmann} {et~al.}(2008){Brinchmann}, {Pettini}, \&
  {Charlot}}]{brinchmann2008}
{Brinchmann}, J., {Pettini}, M., \& {Charlot}, S. 2008, \mnras, 385, 769

\bibitem[{{Calzetti} {et~al.}(2000){Calzetti}, {Armus}, {Bohlin}, {Kinney},
  {Koornneef}, \& {Storchi-Bergmann}}]{calzetti2000}
{Calzetti}, D., {Armus}, L., {Bohlin}, R.~C., {et~al.} 2000, \apj, 533, 682

\bibitem[{{Cardelli} {et~al.}(1989){Cardelli}, {Clayton}, \&
  {Mathis}}]{cardelli1989}
{Cardelli}, J.~A., {Clayton}, G.~C., \& {Mathis}, J.~S. 1989, \apj, 345, 245

\bibitem[{{Chabrier}(2003)}]{chabrier2003}
{Chabrier}, G. 2003, \pasp, 115, 763

\bibitem[{{Coil} {et~al.}(2014){Coil}, {Aird}, {Reddy}, {Shapley}, {Kriek},
  {Siana}, {Mobasher}, {Freeman}, {Price}, \& {Shivaei}}]{coil2014}
{Coil}, A.~L., {Aird}, J., {Reddy}, N.~A., {et~al.} 2014, ArXiv e-prints,
  arXiv:1409.6522

\bibitem[{{Conroy} {et~al.}(2009){Conroy}, {Gunn}, \& {White}}]{conroy2009}
{Conroy}, C., {Gunn}, J.~E., \& {White}, M. 2009, \apj, 699, 486

\bibitem[{{Dav{\'e}} {et~al.}(2011){Dav{\'e}}, {Finlator}, \&
  {Oppenheimer}}]{dave2011}
{Dav{\'e}}, R., {Finlator}, K., \& {Oppenheimer}, B.~D. 2011, \mnras, 416, 1354

\bibitem[{{Dav{\'e}} {et~al.}(2012){Dav{\'e}}, {Finlator}, \&
  {Oppenheimer}}]{dave2012}
---. 2012, \mnras, 421, 98

\bibitem[{{Dom{\'{\i}}nguez} {et~al.}(2013){Dom{\'{\i}}nguez}, {Siana},
  {Henry}, {Scarlata}, {Bedregal}, {Malkan}, {Atek}, {Ross}, {Colbert},
  {Teplitz}, {Rafelski}, {McCarthy}, {Bunker}, {Hathi}, {Dressler}, {Martin},
  \& {Masters}}]{dominguez2013}
{Dom{\'{\i}}nguez}, A., {Siana}, B., {Henry}, A.~L., {et~al.} 2013, \apj, 763,
  145

\bibitem[{{Dopita} {et~al.}(2013){Dopita}, {Sutherland}, {Nicholls}, {Kewley},
  \& {Vogt}}]{dopita2013}
{Dopita}, M.~A., {Sutherland}, R.~S., {Nicholls}, D.~C., {Kewley}, L.~J., \&
  {Vogt}, F.~P.~A. 2013, \apjs, 208, 10

\bibitem[{{Eldridge} \& {Stanway}(2009)}]{eldridge2009}
{Eldridge}, J.~J., \& {Stanway}, E.~R. 2009, \mnras, 400, 1019

\bibitem[{{Erb} {et~al.}(2006{\natexlab{a}}){Erb}, {Shapley}, {Pettini},
  {Steidel}, {Reddy}, \& {Adelberger}}]{erb2006a}
{Erb}, D.~K., {Shapley}, A.~E., {Pettini}, M., {et~al.} 2006{\natexlab{a}},
  \apj, 644, 813

\bibitem[{{Erb} {et~al.}(2006{\natexlab{b}}){Erb}, {Steidel}, {Shapley},
  {Pettini}, {Reddy}, \& {Adelberger}}]{erb2006c}
{Erb}, D.~K., {Steidel}, C.~C., {Shapley}, A.~E., {et~al.} 2006{\natexlab{b}},
  \apj, 647, 128

\bibitem[{{Ferland} {et~al.}(1998){Ferland}, {Korista}, {Verner}, {Ferguson},
  {Kingdon}, \& {Verner}}]{ferland1998}
{Ferland}, G.~J., {Korista}, K.~T., {Verner}, D.~A., {et~al.} 1998, \pasp, 110,
  761

\bibitem[{{Finlator} \& {Dav{\'e}}(2008)}]{finlator2008}
{Finlator}, K., \& {Dav{\'e}}, R. 2008, \mnras, 385, 2181

\bibitem[{{F{\"o}rster Schreiber} {et~al.}(2009){F{\"o}rster Schreiber},
  {Genzel}, {Bouch{\'e}}, {Cresci}, {Davies}, {Buschkamp}, {Shapiro},
  {Tacconi}, {Hicks}, {Genel}, {Shapley}, {Erb}, {Steidel}, {Lutz},
  {Eisenhauer}, {Gillessen}, {Sternberg}, {Renzini}, {Cimatti}, {Daddi},
  {Kurk}, {Lilly}, {Kong}, {Lehnert}, {Nesvadba}, {Verma}, {McCracken},
  {Arimoto}, {Mignoli}, \& {Onodera}}]{forsterschreiber2009}
{F{\"o}rster Schreiber}, N.~M., {Genzel}, R., {Bouch{\'e}}, N., {et~al.} 2009,
  \apj, 706, 1364

\bibitem[{{Grogin} {et~al.}(2011){Grogin}, {Kocevski}, {Faber}, {Ferguson},
  {Koekemoer}, {Riess}, {Acquaviva}, {Alexander}, {Almaini}, {Ashby}, {Barden},
  {Bell}, {Bournaud}, {Brown}, {Caputi}, {Casertano}, {Cassata}, {Castellano},
  {Challis}, {Chary}, {Cheung}, {Cirasuolo}, {Conselice}, {Roshan Cooray},
  {Croton}, {Daddi}, {Dahlen}, {Dav{\'e}}, {de Mello}, {Dekel}, {Dickinson},
  {Dolch}, {Donley}, {Dunlop}, {Dutton}, {Elbaz}, {Fazio}, {Filippenko},
  {Finkelstein}, {Fontana}, {Gardner}, {Garnavich}, {Gawiser}, {Giavalisco},
  {Grazian}, {Guo}, {Hathi}, {H{\"a}ussler}, {Hopkins}, {Huang}, {Huang},
  {Jha}, {Kartaltepe}, {Kirshner}, {Koo}, {Lai}, {Lee}, {Li}, {Lotz}, {Lucas},
  {Madau}, {McCarthy}, {McGrath}, {McIntosh}, {McLure}, {Mobasher},
  {Moustakas}, {Mozena}, {Nandra}, {Newman}, {Niemi}, {Noeske}, {Papovich},
  {Pentericci}, {Pope}, {Primack}, {Rajan}, {Ravindranath}, {Reddy}, {Renzini},
  {Rix}, {Robaina}, {Rodney}, {Rosario}, {Rosati}, {Salimbeni}, {Scarlata},
  {Siana}, {Simard}, {Smidt}, {Somerville}, {Spinrad}, {Straughn}, {Strolger},
  {Telford}, {Teplitz}, {Trump}, {van der Wel}, {Villforth}, {Wechsler},
  {Weiner}, {Wiklind}, {Wild}, {Wilson}, {Wuyts}, {Yan}, \& {Yun}}]{grogin2011}
{Grogin}, N.~A., {Kocevski}, D.~D., {Faber}, S.~M., {et~al.} 2011, \apjs, 197,
  35

\bibitem[{{Hainline} {et~al.}(2009){Hainline}, {Shapley}, {Kornei}, {Pettini},
  {Buckley-Geer}, {Allam}, \& {Tucker}}]{hainline2009}
{Hainline}, K.~N., {Shapley}, A.~E., {Kornei}, K.~A., {et~al.} 2009, \apj, 701,
  52

\bibitem[{{Juneau} {et~al.}(2014){Juneau}, {Bournaud}, {Charlot}, {Daddi},
  {Elbaz}, {Trump}, {Brinchmann}, {Dickinson}, {Duc}, {Gobat}, {Jean-Baptiste},
  {Le Floc'h}, {Lehnert}, {Pacifici}, {Pannella}, \& {Schreiber}}]{juneau2014}
{Juneau}, S., {Bournaud}, F., {Charlot}, S., {et~al.} 2014, \apj, 788, 88

\bibitem[{{Kauffmann} {et~al.}(2003){Kauffmann}, {Heckman}, {Tremonti},
  {Brinchmann}, {Charlot}, {White}, {Ridgway}, {Brinkmann}, {Fukugita}, {Hall},
  {Ivezi{\' c}}, {Richards}, \& {Schneider}}]{kauffmann2003}
{Kauffmann}, G., {Heckman}, T.~M., {Tremonti}, C., {et~al.} 2003, \mnras, 346,
  1055

\bibitem[{{Kennicutt}(1998)}]{kennicutt1998}
{Kennicutt}, R.~C. 1998, \araa, 36, 189

\bibitem[{{Kewley} \& {Dopita}(2002)}]{kewley2002}
{Kewley}, L.~J., \& {Dopita}, M.~A. 2002, \apjs, 142, 35

\bibitem[{{Kewley} {et~al.}(2013){Kewley}, {Dopita}, {Leitherer}, {Dav{\'e}},
  {Yuan}, {Allen}, {Groves}, \& {Sutherland}}]{kewley2013}
{Kewley}, L.~J., {Dopita}, M.~A., {Leitherer}, C., {et~al.} 2013, \apj, 774,
  100

\bibitem[{{Kewley} {et~al.}(2001){Kewley}, {Dopita}, {Sutherland}, {Heisler},
  \& {Trevena}}]{kewley2001}
{Kewley}, L.~J., {Dopita}, M.~A., {Sutherland}, R.~S., {Heisler}, C.~A., \&
  {Trevena}, J. 2001, \apj, 556, 121

\bibitem[{{Kewley} \& {Ellison}(2008)}]{kewley2008}
{Kewley}, L.~J., \& {Ellison}, S.~L. 2008, \apj, 681, 1183

\bibitem[{{Koekemoer} {et~al.}(2011){Koekemoer}, {Faber}, {Ferguson}, {Grogin},
  {Kocevski}, {Koo}, {Lai}, {Lotz}, {Lucas}, {McGrath}, {Ogaz}, {Rajan},
  {Riess}, {Rodney}, {Strolger}, {Casertano}, {Castellano}, {Dahlen},
  {Dickinson}, {Dolch}, {Fontana}, {Giavalisco}, {Grazian}, {Guo}, {Hathi},
  {Huang}, {van der Wel}, {Yan}, {Acquaviva}, {Alexander}, {Almaini}, {Ashby},
  {Barden}, {Bell}, {Bournaud}, {Brown}, {Caputi}, {Cassata}, {Challis},
  {Chary}, {Cheung}, {Cirasuolo}, {Conselice}, {Roshan Cooray}, {Croton},
  {Daddi}, {Dav{\'e}}, {de Mello}, {de Ravel}, {Dekel}, {Donley}, {Dunlop},
  {Dutton}, {Elbaz}, {Fazio}, {Filippenko}, {Finkelstein}, {Frazer}, {Gardner},
  {Garnavich}, {Gawiser}, {Gruetzbauch}, {Hartley}, {H{\"a}ussler},
  {Herrington}, {Hopkins}, {Huang}, {Jha}, {Johnson}, {Kartaltepe},
  {Khostovan}, {Kirshner}, {Lani}, {Lee}, {Li}, {Madau}, {McCarthy},
  {McIntosh}, {McLure}, {McPartland}, {Mobasher}, {Moreira}, {Mortlock},
  {Moustakas}, {Mozena}, {Nandra}, {Newman}, {Nielsen}, {Niemi}, {Noeske},
  {Papovich}, {Pentericci}, {Pope}, {Primack}, {Ravindranath}, {Reddy},
  {Renzini}, {Rix}, {Robaina}, {Rosario}, {Rosati}, {Salimbeni}, {Scarlata},
  {Siana}, {Simard}, {Smidt}, {Snyder}, {Somerville}, {Spinrad}, {Straughn},
  {Telford}, {Teplitz}, {Trump}, {Vargas}, {Villforth}, {Wagner}, {Wandro},
  {Wechsler}, {Weiner}, {Wiklind}, {Wild}, {Wilson}, {Wuyts}, \&
  {Yun}}]{koekemoer2011}
{Koekemoer}, A.~M., {Faber}, S.~M., {Ferguson}, H.~C., {et~al.} 2011, \apjs,
  197, 36

\bibitem[{{Kriek} {et~al.}(2009){Kriek}, {van Dokkum}, {Labb{\'e}}, {Franx},
  {Illingworth}, {Marchesini}, \& {Quadri}}]{kriek2009}
{Kriek}, M., {van Dokkum}, P.~G., {Labb{\'e}}, I., {et~al.} 2009, \apj, 700,
  221

\bibitem[{{Kriek} {et~al.}(2014){Kriek}, {Shapley}, {Reddy}, {Siana}, {Coil},
  {Mobasher}, {Freeman}, {de Groot}, {Price}, {Sanders}, {Shivaei}, {Brammer},
  {Momcheva}, {Skelton}, {van Dokkum}, {Whitaker}, {Aird}, {Azadi}, {Kassis},
  {Bullock}, {Conroy}, {Dave}, {Keres}, \& {Krumholz}}]{kriek2014}
{Kriek}, M., {Shapley}, A.~E., {Reddy}, N.~A., {et~al.} 2014, ArXiv e-prints,
  arXiv:1412.1835

\bibitem[{{Lilly} {et~al.}(2003){Lilly}, {Carollo}, \& {Stockton}}]{lilly2003}
{Lilly}, S.~J., {Carollo}, C.~M., \& {Stockton}, A.~N. 2003, \apj, 597, 730

\bibitem[{{Liu} {et~al.}(2008){Liu}, {Shapley}, {Coil}, {Brinchmann}, \&
  {Ma}}]{liu2008}
{Liu}, X., {Shapley}, A.~E., {Coil}, A.~L., {Brinchmann}, J., \& {Ma}, C.-P.
  2008, \apj, 678, 758

\bibitem[{{Maiolino} {et~al.}(2008){Maiolino}, {Nagao}, {Grazian}, {Cocchia},
  {Marconi}, {Mannucci}, {Cimatti}, {Pipino}, {Ballero}, {Calura}, {Chiappini},
  {Fontana}, {Granato}, {Matteucci}, {Pastorini}, {Pentericci}, {Risaliti},
  {Salvati}, \& {Silva}}]{maiolino2008}
{Maiolino}, R., {Nagao}, T., {Grazian}, A., {et~al.} 2008, \aap, 488, 463

\bibitem[{{Masters} {et~al.}(2014){Masters}, {McCarthy}, {Siana}, {Malkan},
  {Mobasher}, {Atek}, {Henry}, {Martin}, {Rafelski}, {Hathi}, {Scarlata},
  {Ross}, {Bunker}, {Blanc}, {Bedregal}, {Dom{\'{\i}}nguez}, {Colbert},
  {Teplitz}, \& {Dressler}}]{masters2014}
{Masters}, D., {McCarthy}, P., {Siana}, B., {et~al.} 2014, \apj, 785, 153

\bibitem[{{McLean} {et~al.}(2012){McLean}, {Steidel}, {Epps}, {Konidaris},
  {Matthews}, {Adkins}, {Aliado}, {Brims}, {Canfield}, {Cromer}, {Fucik},
  {Kulas}, {Mace}, {Magnone}, {Rodriguez}, {Rudie}, {Trainor}, {Wang}, {Weber},
  \& {Weiss}}]{mclean2012}
{McLean}, I.~S., {Steidel}, C.~C., {Epps}, H.~W., {et~al.} 2012, in Society of
  Photo-Optical Instrumentation Engineers (SPIE) Conference Series, Vol. 8446,
  Society of Photo-Optical Instrumentation Engineers (SPIE) Conference Series

\bibitem[{{Nagao} {et~al.}(2006){Nagao}, {Maiolino}, \& {Marconi}}]{nagao2006}
{Nagao}, T., {Maiolino}, R., \& {Marconi}, A. 2006, \aap, 459, 85

\bibitem[{{Nakajima} \& {Ouchi}(2014)}]{nakajima2014}
{Nakajima}, K., \& {Ouchi}, M. 2014, \mnras, 442, 900

\bibitem[{{Nakajima} {et~al.}(2013){Nakajima}, {Ouchi}, {Shimasaku},
  {Hashimoto}, {Ono}, \& {Lee}}]{nakajima2013}
{Nakajima}, K., {Ouchi}, M., {Shimasaku}, K., {et~al.} 2013, \apj, 769, 3

\bibitem[{{Newman} {et~al.}(2014){Newman}, {Buschkamp}, {Genzel}, {F{\"o}rster
  Schreiber}, {Kurk}, {Sternberg}, {Gnat}, {Rosario}, {Mancini}, {Lilly},
  {Renzini}, {Burkert}, {Carollo}, {Cresci}, {Davies}, {Eisenhauer}, {Genel},
  {Shapiro Griffin}, {Hicks}, {Lutz}, {Naab}, {Peng}, {Tacconi}, {Wuyts},
  {Zamorani}, {Vergani}, \& {Weiner}}]{newman2014}
{Newman}, S.~F., {Buschkamp}, P., {Genzel}, R., {et~al.} 2014, \apj, 781, 21

\bibitem[{{Osterbrock}(1989)}]{osterbrock1989}
{Osterbrock}, D.~E. 1989, {Astrophysics of gaseous nebulae and active galactic
  nuclei}

\bibitem[{{P{\'e}rez-Montero} \& {Contini}(2009)}]{perezmontero2009}
{P{\'e}rez-Montero}, E., \& {Contini}, T. 2009, \mnras, 398, 949

\bibitem[{{Pettini} \& {Pagel}(2004)}]{pp2004}
{Pettini}, M., \& {Pagel}, B.~E.~J. 2004, \mnras, 348, L59

\bibitem[{{Price} {et~al.}(2014){Price}, {Kriek}, {Brammer}, {Conroy},
  {F{\"o}rster Schreiber}, {Franx}, {Fumagalli}, {Lundgren}, {Momcheva},
  {Nelson}, {Skelton}, {van Dokkum}, {Whitaker}, \& {Wuyts}}]{price2014}
{Price}, S.~H., {Kriek}, M., {Brammer}, G.~B., {et~al.} 2014, \apj, 788, 86

\bibitem[{{Sanders} {et~al.}(2014){Sanders}, {Shapley}, {Kriek}, {Reddy},
  {Freeman}, {Coil}, {Siana}, {Mobasher}, {Shivaei}, {Price}, \& {de
  Groot}}]{sanders2014}
{Sanders}, R.~L., {Shapley}, A.~E., {Kriek}, M., {et~al.} 2014, ArXiv e-prints,
  arXiv:1408.2521

\bibitem[{{S{\'e}rsic}(1968)}]{sersic1968}
{S{\'e}rsic}, J.~L. 1968, {Atlas de galaxias australes}

\bibitem[{{Shapley} {et~al.}(2005){Shapley}, {Coil}, {Ma}, \&
  {Bundy}}]{shapley2005b}
{Shapley}, A.~E., {Coil}, A.~L., {Ma}, C.-P., \& {Bundy}, K. 2005, \apj, 635,
  1006

\bibitem[{{Shapley} {et~al.}(2003){Shapley}, {Steidel}, {Pettini}, \&
  {Adelberger}}]{shapley2003}
{Shapley}, A.~E., {Steidel}, C.~C., {Pettini}, M., \& {Adelberger}, K.~L. 2003,
  \apj, 588, 65

\bibitem[{{Shirazi} {et~al.}(2014){Shirazi}, {Brinchmann}, \&
  {Rahmati}}]{shirazi2014}
{Shirazi}, M., {Brinchmann}, J., \& {Rahmati}, A. 2014, \apj, 787, 120

\bibitem[{{Skelton} {et~al.}(2014){Skelton}, {Whitaker}, {Momcheva}, {Brammer},
  {van Dokkum}, {Labbe}, {Franx}, {van der Wel}, {Bezanson}, {Da Cunha},
  {Fumagalli}, {Foerster Schreiber}, {Kriek}, {Leja}, {Lundgren}, {Magee},
  {Marchesini}, {Maseda}, {Nelson}, {Oesch}, {Pacifici}, {Patel}, {Price},
  {Rix}, {Tal}, {Wake}, \& {Wuyts}}]{skelton2014}
{Skelton}, R.~E., {Whitaker}, K.~E., {Momcheva}, I.~G., {et~al.} 2014, ArXiv
  e-prints, arXiv:1403.3689

\bibitem[{{Steidel} {et~al.}(2014){Steidel}, {Rudie}, {Strom}, {Pettini},
  {Reddy}, {Shapley}, {Trainor}, {Erb}, {Turner}, {Konidaris}, {Kulas}, {Mace},
  {Matthews}, \& {McLean}}]{steidel2014}
{Steidel}, C.~C., {Rudie}, G.~C., {Strom}, A.~L., {et~al.} 2014, ArXiv
  e-prints, arXiv:1405.5473

\bibitem[{{Storey} \& {Zeippen}(2000)}]{storey2000}
{Storey}, P.~J., \& {Zeippen}, C.~J. 2000, \mnras, 312, 813

\bibitem[{{Tremonti} {et~al.}(2004){Tremonti}, {Heckman}, {Kauffmann},
  {Brinchmann}, {Charlot}, {White}, {Seibert}, {Peng}, {Schlegel}, {Uomoto},
  {Fukugita}, \& {Brinkmann}}]{tremonti2004}
{Tremonti}, C.~A., {Heckman}, T.~M., {Kauffmann}, G., {et~al.} 2004, \apj, 613,
  898

\bibitem[{{Veilleux} \& {Osterbrock}(1987)}]{veilleux1987}
{Veilleux}, S., \& {Osterbrock}, D.~E. 1987, \apjs, 63, 295

\bibitem[{{Verdolini} {et~al.}(2013){Verdolini}, {Yeh}, {Krumholz}, {Matzner},
  \& {Tielens}}]{verdolini2013}
{Verdolini}, S., {Yeh}, S.~C.~C., {Krumholz}, M.~R., {Matzner}, C.~D., \&
  {Tielens}, A.~G.~G.~M. 2013, \apj, 769, 12

\bibitem[{{Wright} {et~al.}(2010){Wright}, {Larkin}, {Graham}, \&
  {Ma}}]{wright2010}
{Wright}, S.~A., {Larkin}, J.~E., {Graham}, J.~R., \& {Ma}, C.-P. 2010, \apj,
  711, 1291

\bibitem[{{Yeh} {et~al.}(2013){Yeh}, {Verdolini}, {Krumholz}, {Matzner}, \&
  {Tielens}}]{yeh2013}
{Yeh}, S.~C.~C., {Verdolini}, S., {Krumholz}, M.~R., {Matzner}, C.~D., \&
  {Tielens}, A.~G.~G.~M. 2013, \apj, 769, 11

\bibitem[{{York} {et~al.}(2000){York}, {Adelman}, {Anderson}, {Anderson},
  {Annis}, {Bahcall}, {Bakken}, {Barkhouser}, {Bastian}, {Berman}, {Boroski},
  {Bracker}, {Briegel}, {Briggs}, {Brinkmann}, {Brunner}, {Burles}, {Carey},
  {Carr}, {Castander}, {Chen}, {Colestock}, {Connolly}, {Crocker}, {Csabai},
  {Czarapata}, {Davis}, {Doi}, {Dombeck}, {Eisenstein}, {Ellman}, {Elms},
  {Evans}, {Fan}, {Federwitz}, {Fiscelli}, {Friedman}, {Frieman}, {Fukugita},
  {Gillespie}, {Gunn}, {Gurbani}, {de Haas}, {Haldeman}, {Harris}, {Hayes},
  {Heckman}, {Hennessy}, {Hindsley}, {Holm}, {Holmgren}, {Huang}, {Hull},
  {Husby}, {Ichikawa}, {Ichikawa}, {Ivezi{\'c}}, {Kent}, {Kim}, {Kinney},
  {Klaene}, {Kleinman}, {Kleinman}, {Knapp}, {Korienek}, {Kron}, {Kunszt},
  {Lamb}, {Lee}, {Leger}, {Limmongkol}, {Lindenmeyer}, {Long}, {Loomis},
  {Loveday}, {Lucinio}, {Lupton}, {MacKinnon}, {Mannery}, {Mantsch}, {Margon},
  {McGehee}, {McKay}, {Meiksin}, {Merelli}, {Monet}, {Munn}, {Narayanan},
  {Nash}, {Neilsen}, {Neswold}, {Newberg}, {Nichol}, {Nicinski}, {Nonino},
  {Okada}, {Okamura}, {Ostriker}, {Owen}, {Pauls}, {Peoples}, {Peterson},
  {Petravick}, {Pier}, {Pope}, {Pordes}, {Prosapio}, {Rechenmacher}, {Quinn},
  {Richards}, {Richmond}, {Rivetta}, {Rockosi}, {Ruthmansdorfer}, {Sandford},
  {Schlegel}, {Schneider}, {Sekiguchi}, {Sergey}, {Shimasaku}, {Siegmund},
  {Smee}, {Smith}, {Snedden}, {Stone}, {Stoughton}, {Strauss}, {Stubbs},
  {SubbaRao}, {Szalay}, {Szapudi}, {Szokoly}, {Thakar}, {Tremonti}, {Tucker},
  {Uomoto}, {Vanden Berk}, {Vogeley}, {Waddell}, {Wang}, {Watanabe},
  {Weinberg}, {Yanny}, {Yasuda}, \& {SDSS Collaboration}}]{york2000}
{York}, D.~G., {Adelman}, J., {Anderson}, Jr., J.~E., {et~al.} 2000, \aj, 120,
  1579

\end{thebibliography}

\end{document}